\documentclass[twocolumn,showpacs,pra,superscriptaddress,floatfix]{revtex4}

\usepackage{color}
\usepackage{tabularx}
\usepackage{epsfig}
\usepackage{amsmath}
\usepackage{amssymb}
\usepackage{graphicx}
\usepackage{wasysym}

\begin{document}

\title{Topological color codes on Union Jack lattices:\\
a stable implementation of the whole Clifford group}

\author{Helmut G.~Katzgraber}
\affiliation {Department of Physics and Astronomy, Texas A\&M
University, College Station, Texas 77843-4242, USA}
\affiliation {Theoretische Physik, ETH Zurich, CH-8093 Zurich, Switzerland}

\author{H.~Bombin}
\affiliation{Perimeter Institute for Theoretical Physics,
Waterloo, Ontario N2L 2Y5, Canada}

\author{Ruben S.~Andrist}
\affiliation {Theoretische Physik, ETH Zurich, CH-8093 Zurich, Switzerland}

\author{M.~A.~Martin-Delgado}
\affiliation{Departamento de F{\'i}sica Te{\'o}rica I, Universidad
Complutense, 28040 Madrid, Spain}


\begin{abstract}

We study the error threshold of topological color codes on Union Jack
lattices that allow for the full implementation of the whole Clifford
group of quantum gates. After mapping the error-correction process onto
a statistical mechanical random three-body Ising model on a Union Jack
lattice, we compute its phase diagram in the temperature--disorder
plane using Monte Carlo simulations. Surprisingly, topological color
codes on Union Jack lattices have a similar error stability to color
codes on triangular lattices, as well as to the Kitaev toric code. The
enhanced computational capabilities of the topological color codes on
Union Jack lattices with respect to triangular lattices and the toric
code combined with the inherent robustness of this implementation
show good prospects for future stable quantum computer implementations.

\end{abstract}



\maketitle

\section{Introduction}
\label{sec:introduction}

Recently, the error threshold for topological color codes
on triangular lattices has been computed using Monte Carlo
methods \cite{katzgraber:09c}, as well as using duality arguments
\cite{ohzeki:09a}. Understanding the error-correction properties of
topological color codes is of paramount importance and serves as
a benchmark for comparing to other quantum error-correction codes.
Being one of the central figures of merit, error thresholds describe
the tolerable value of noise below which a quantum code can perform
quantum operations without the impact of decoherence effects due to
the environment.

In particular, for topological stabilizer codes there are two main
instances of quantum codes that are simple enough such that detailed
studies of their computational capabilities can be carried out: the
original toric code (TC) introduced by Kitaev \cite{kitaev:03} and
topological color codes (TCC) \cite{bombin:06}. In both cases, quantum
gates are implemented by external operations that act nontrivially
on the degenerate ground-state manifold of a given quantum lattice
Hamiltonian. The ground state harbors the topological stabilizer code
while excited states correspond to errors in the system.  Within this
scheme, topological quantum operations are performed based on 
ground-state properties \cite{bombin:07b}.  Alternatively, it is possible to
implement topological quantum gates based on quasiparticle braiding
operations \cite{nayak:08}.

In comparison to the toric code, TCCs can encode a larger number
of qubits at a given surface of fixed topology and the variety
of topological quantum gates that can be applied transversally is
larger \cite{bombin:07c}.  In particular, it is the possibility to
perform the whole Clifford group of quantum gates in a transversal
and topological way that makes TCCs especially appealing in quantum
information theory.  The error threshold is one figure of merit for
the performance of any topological stabilizer code.  If this value is
very small for a given code then all the advantages of the new code
are meaningless in practical applications because small amounts of
external noise would spoil its stability.

Improving the computational capabilities of a quantum information
system generally comes at the price of decreased robustness against
environmental noise. However, the exact threshold value for the
qubit error rate $p$ in TCCs on triangular lattices has been
computed numerically to be $p_{\rm c}=0.109(2)$ \cite{katzgraber:09c},
which agrees within error bars with the value for the toric code
\cite{honecker:01,merz:02,ohzeki:09,parisen:09}. This result---also
confirmed by other studies \cite{ohzeki:09a,wang:09,landahl:09}---is
encouraging: Topological color codes preserve a high error tolerance
for quantum operations without performance trade-offs (i.e., it
is possible to perform complex quantum operations such as quantum
distillation, teleportation, and dense coding while retaining the
same stability against noise).

To estimate the error tolerance of a TCC, the error-correction process
is mapped onto a statistical model with random three-body interactions
that correspond to the faulty bits. In analogy to the Kitaev model, the
random three-body Ising model \cite{katzgraber:09c} plays a similar role as
the random bond Ising model in the Kitaev toric code \cite{dennis:02}.
The study of the three-body random Ising model further highlights the
relationship between spin-glass physics and quantum information theory,
and presents a new class of model systems exhibiting glassy behavior
via three-body interactions (i.e., without spin-reversal symmetry).

In this work we revisit the problem and estimate the error threshold
for topological color codes on the Union Jack (UJ) lattice. The
motivation to use the triangular lattice (which is dual to the
hexagonal lattice) in Ref.~\cite{katzgraber:09c} was based on the
triangular lattice being the simplest example of a family of lattices
named colexes (color complexes) \cite{bombin:07} and the simplicity
of numerical simulations.  However, there is a technical caveat
with TCCs on the hexagonal lattice: they do not fully reproduce the
Clifford group \cite{bombin:06}.  In fact, one of the gate generators
of the group needs the square-octagonal lattice for its topological
implementation. We therefore use the fact that the UJ lattice is the
dual lattice of the square-octagonal lattice.

There is another fundamental reason to study TCCs on the UJ lattice.
The presence of three-body random interactions poses new problems that
were absent in the Kitaev code. For example, the type of lattice and
its connectivity may play a role in some of the essential features
of the mapped statistical model that is visible in the location of the
multicritical point or the values of the critical exponents.  To see
this, there is another route to establish connections between quantum
information and classical statistical mechanical models. It is based
on the notion of classical simulability of topological ground states
\cite{bravyi:07,bombin:08} and classical models with the completeness
property \cite{vandennest:08,delascuevas:09}.  The implication of
these studies for TCCs is that certain properties of their ground
state can be related to a three-body Ising model without randomness ($p=0$)
\cite{bombin:08}. Following universality \cite{yeomans:92,cardy:96},
it is expected that the critical exponents depend on the lattice
geometry and order parameter symmetry \cite{hintermann:72,baxter:73}.
Therefore, TCCs allow us to see a connection between critical
exponents and computational capabilities of a quantum code, and not
just the location of its critical point. This is a novel feature that
is absent in the statistical mapping from topological codes and random
models \cite{dennis:02} where only the location of the multicritical
point plays a role in the features of the quantum information system.
We conjecture that the same connection between lattice-dependent
critical exponents and different quantum capabilities for TCCs holds
in the presence of randomness $p\neq 0$ at the multicritical point.

Our simulations show that the numerical value of the multicritical
point $p_{\rm c}$ for the three-body Ising model on random UJ lattices agrees
within error bars with the value obtained for the triangular lattice
\cite{katzgraber:09c}. In turn, this  highlights the stability of TCCs
to external noise. In addition, we compare the phase boundary between
a ferromagnetic ground state and a paramagnetic one for the three-body
Ising model on both triangular (TR) and UJ lattices. Our results show
that these are rather similar.

Note that throughout this article we assume that
external error correction is carried out on the
protected system represented by the topological color
code. This is the standard notion of quantum error correction
\cite{shor:95,steane:96,calderbank:96,steane:96a,gottesman:96,bombin:07d}.
Recently, experimental realizations of topological error correction
have been implemented \cite{gao:09}, as well as experimental proposals
for TCCs using Rydberg atoms \cite{weimer:09} have been made.  However,
there are also more demanding schemes in which the topological
color code can be internally protected leading to the notion of a
self-correcting quantum computer \cite{bombin:09a,alicki:08}.

The paper is organized as follows: In Sec.~\ref{sec:ccodes} we
introduce topological color-code states on Union Jack lattices,
followed by the mapping of the error-correction process of the
TCC onto random three-body Ising models in Sec.~\ref{sec:mapping}.
In Sec.~\ref{sec:numerics} we describe our Monte Carlo simulations
and how to locate the multicritical point $p_{\rm c}$ that corresponds to
the error threshold of the TCCs.  Results from simulations are shown
in Sec.~\ref{sec:results}, followed by concluding remarks.

\section{Color codes on the Union Jack lattice}
\label{sec:ccodes}

\begin{figure}

\includegraphics[width=0.75\columnwidth]{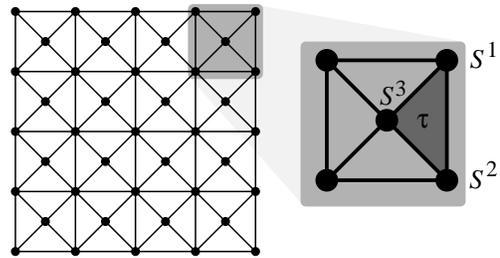}

\caption{
Simulated Union Jack lattice: The sites at the edges of the squares
(inset) have coordination $8$, whereas the sites in the center
of the square plaquettes have coordination $4$. The Hamiltonian in
Eq.~(\ref{eq:ham}) is a sum over all possible triangles with a weight
$\tau$ multiplied by the product of the three spins $S^1$, $S^2$,
and $S^3$ on the triangle's edges (inset).  For the simulations, we use
periodic boundary conditions (i.e., the lattice is placed on a torus).
}
\label{fig:lattice}
\end{figure}

A TCC can be obtained from any two-dimensional (2D) lattice in
which all plaquettes are triangles and vertices are three-colorable
such that no link connects vertices of the same color. It is
also possible to work in the dual lattice (called a 2-colex, see
Refs.~\cite{bombin:06,bombin:07}) but here we prefer to work in
the triangular lattice to have a more direct mapping, as done in
Ref.~\cite{katzgraber:09c}. The lattice is embedded in a compact
surface of arbitrary topology. Since information is encoded in
topological degrees of freedom, the code is nontrivial only when the
topology of the surface is nontrivial (i.e., the genus $g$ of the
surface has to be $g \ge 1$). Note that data can also be encoded in a
planar surface with holes and appropriate boundary conditions. This
makes topological proposals more amenable to experimental setups
\cite{bravyi:98,bombin:09}.  In this work we embed the system in
a Union Jack lattice (see Fig.~\ref{fig:lattice}) to be able to
transversally implement all Clifford gates  \cite{bombin:06}.

The construction of the code $\mathcal C$ is done in the following way:
Consider a physical system with a qubit at each lattice triangle and
introduce the following vertex operators that generate the stabilizer
group of $\mathcal C$ \cite{gottesman:96}.  For each vertex $v$ there
are two types of operators that correspond to Pauli operators of $X$
or $Z$ type:
\begin{equation}
X_v :=\bigotimes_{\triangle : v \in \triangle}
X_{\triangle}, \qquad Z_v :=\bigotimes_{\triangle : v \in \triangle}
Z_{\triangle}.
\label{eq:vertexOps}
\end{equation}
that is, a vertex operator acts on all nearby triangles which may
be 4 or 8 (see Fig.~\ref{fig:lattice}). Vertex operators pairwise
commute and square to identity, so that they generate an Abelian
group $\mathcal S$ called the stabilizer. The code $\mathcal C$ is
defined as the subspace stabilized by $\mathcal S$ and thus contains
the states with $X_v=Z_v=1$ $\forall v$.

What makes the UJ lattice special is that the vertex operators---the
generators of the stabilizer---have support on a number of qubits
that is a multiple of four. This fact, together with the properties of
general color codes, gives rise to an important feature, namely that
Clifford gates leave the stabilizer invariant. After the introduction
of suitable boundaries on the lattice, this allows for the transversal
implementation of the Clifford group on color codes \cite{bombin:06}.

\section{Error-Correction Mapping}
\label{sec:mapping}

When an encoded state in a color code $\mathcal C$ is subject
to errors, the first step for their possible correction is the
measurement of vertex operators. The resulting collection of $\pm 1$
eigenvalues is called the error syndrome and gives information about
which errors occurred. Indeed, the errors need not be guessed exactly
but only up to a stabilizer element.

Color codes have a structure with stabilizer generators that are
either products of $X$ or $Z$ Pauli operators, but not both. This
allows us to treat bit-flip and phase errors separately: $X$-type
($Z$-type) errors produce violations of $Z$-type ($X$-type) vertex
operators. The correction of each type of error can be expressed
in homological terms (see Ref.~\cite{katzgraber:09c}) such that the
syndrome gives the boundary of the error. In order to do a successful
correction, errors must be guessed only up to homology. In contrast
to toric codes where one deals with the usual homology of paths on
a surface, in TCCs two different types (colors) of paths are allowed.

Consider a standard error model based on stochastic errors in which
phase errors $Z$ and bit-flip errors $X$ are uncorrelated and occur
with probability $p$ at each qubit. In topological codes, when $p$
is below a threshold $p_{\rm c}$, error correction can be performed with
perfect accuracy in the limit of infinite system size. Above the
threshold, error correction becomes useless in the same limit. In
Ref.~\cite{dennis:02} this error threshold was computed for toric
codes by mapping the error-correction process to a random bond Ising
model.  The corresponding mapping for color codes was carried out
in Ref.~\cite{katzgraber:09c}, which showed that a random three-body Ising
model appears. Note that other general extensions of statistical
mappings are also possible in the context of topological subsystem
codes \cite{bombin:09b}.

Let us summarize what happens when we apply the mapping to the Union
Jack lattice. First, the spins of the statistical model correspond
to the stabilizer generators. Thus, we have a classical spin at
each vertex of the lattice. As for the three-body interactions, they
correspond to the physical qubits and hence there is such a term
per triangle. Each of these terms carries a random sign yielding
the Hamiltonian
\begin{equation}
{\mathcal H} =  -J \sum_{\triangle} \tau_{\triangle}
                 S_\triangle^1
                 S_\triangle^2
                 S_\triangle^3 \; ,
\label{eq:ham}
\end{equation}
where the coupling $J$ is positive and the sum is over the product
of the spins at the vertices of all triangles on the UJ lattice and
$\tau_\triangle = \pm 1$ \cite{comment:J}. The (classical) Ising
spins $S_\triangle^i$ can have the values $\pm 1$. The sign of the
coupling constants $\tau$ are independent random quenched variables,
that are negative with probability $p$.

At low temperature $T$ and small disorder $p$, the system is
ferromagnetic. Above a critical line $p_{\rm crit}(T)$, the model
undergoes a phase transition to a paramagnetic state. The error
threshold $p_{\rm c}$ is given by the crossing point between $p_{\rm
crit}(T)$ and the Nishimori line \cite{nishimori:81,nishimori:01}
\begin{equation}
\exp(-2J/T)=\frac p{(1-p)}.
\end{equation}

\begin{table}[!tb]
\caption{
Simulation parameters: $L$ is the system size, $N_{\rm sa}$ is the
number of disorder samples, $t_{\rm eq} = 2^{b}$ is the number of
equilibration sweeps, $T_{\rm min}$ [$T_{\rm max}$] is the lowest
[highest] temperature, and $N_{\rm T}$ the number of temperatures used.
\label{tab:simparams}}
{\footnotesize
\begin{tabular*}{\columnwidth}{@{\extracolsep{\fill}} l r r r r r r}
\hline
\hline
$p$ & $L$ & $N_{\rm sa}$ & $b$ & $T_{\rm min}$ & $T_{\rm max}$ &$N_{\rm T}$ \\
\hline
$0.00$            & $12$, $18$ & $50$      & $18$ & $2.200$ & $2.350$ & $31$\\
$0.00$            & $24$, $30$ & $50$      & $19$ & $2.200$ & $2.350$ & $31$\\
$0.00$            &       $36$ & $50$      & $20$ & $2.200$ & $2.350$ & $31$\\
$0.01$            & $12$, $18$ & $5\,000$  & $18$ & $1.900$ & $2.400$ & $51$\\
$0.01$            & $24$, $30$ & $5\,000$  & $19$ & $1.900$ & $2.400$ & $51$\\
$0.01$            &       $36$ & $5\,000$  & $20$ & $1.900$ & $2.400$ & $51$\\
$0.02$            & $12$, $18$ & $5\,000$  & $18$ & $1.900$ & $2.400$ & $51$\\
$0.02$            & $24$, $30$ & $5\,000$  & $19$ & $1.900$ & $2.400$ & $51$\\
$0.02$            &       $36$ & $5\,000$  & $20$ & $1.900$ & $2.400$ & $51$\\
$0.03$            & $12$, $18$ & $5\,000$  & $18$ & $1.700$ & $2.200$ & $51$\\
$0.03$            & $24$, $30$ & $5\,000$  & $19$ & $1.700$ & $2.200$ & $51$\\
$0.03$            &       $36$ & $5\,000$  & $20$ & $1.700$ & $2.200$ & $51$\\
$0.04$            & $12$, $18$ & $5\,000$  & $18$ & $1.700$ & $2.200$ & $51$\\
$0.04$            & $24$, $30$ & $5\,000$  & $19$ & $1.700$ & $2.200$ & $51$\\
$0.04$            &       $36$ & $5\,000$  & $20$ & $1.700$ & $2.200$ & $51$\\
$0.06$            & $12$, $18$ & $5\,000$  & $18$ & $1.600$ & $2.100$ & $51$\\
$0.06$            & $24$, $30$ & $5\,000$  & $19$ & $1.600$ & $2.100$ & $51$\\
$0.06$            &       $36$ & $5\,000$  & $20$ & $1.600$ & $2.100$ & $51$\\
$0.08$            & $12$, $18$ & $5\,000$  & $18$ & $1.400$ & $2.000$ & $61$\\
$0.08$            & $24$, $30$ & $5\,000$  & $19$ & $1.400$ & $2.000$ & $61$\\
$0.08$            &       $36$ & $5\,000$  & $20$ & $1.400$ & $2.000$ & $61$\\
$0.10$ --- $0.11$ & $12$, $18$ & $10\,000$ & $18$ & $0.750$ & $2.600$ & $38$\\
$0.10$ --- $0.11$ & $24$, $30$ & $10\,000$ & $19$ & $0.750$ & $2.600$ & $38$\\
$0.10$ --- $0.11$ &       $36$ & $10\,000$ & $20$ & $0.750$ & $2.600$ & $38$\\
$0.10$ --- $0.11$ &       $42$ & $10\,000$ & $22$ & $0.750$ & $2.600$ & $38$\\
$0.12$            & $12$, $18$ & $5\,000$  & $18$ & $0.750$ & $2.600$ & $38$\\
$0.12$            & $24$, $30$ & $5\,000$  & $19$ & $0.750$ & $2.600$ & $38$\\
$0.12$            &       $36$ & $5\,000$  & $20$ & $0.750$ & $2.600$ & $38$\\
\hline
\hline
\end{tabular*}
}
\end{table}

\section{Numerical details}
\label{sec:numerics}

When studying phase transitions of statistical systems, it is
most favorable to study quantities whose finite-size scaling form
is dimensionless (i.e., there is not a system-size dependent
prefactor in front of the scaling function). Typical quantities
are the Binder ratio \cite{binder:86} as well as the dimensionless
two-point finite-size correlation length divided by the system size
\cite{palassini:99b}. The latter has been shown to be a more robust
measure of transition temperatures for systems without spin-reversal
symmetry \cite{katzgraber:09c} because the Binder ratio becomes
negative and steep at the transition (i.e., pinpointing the critical
temperature is difficult).

\begin{figure*}

\includegraphics[width=6.0cm]{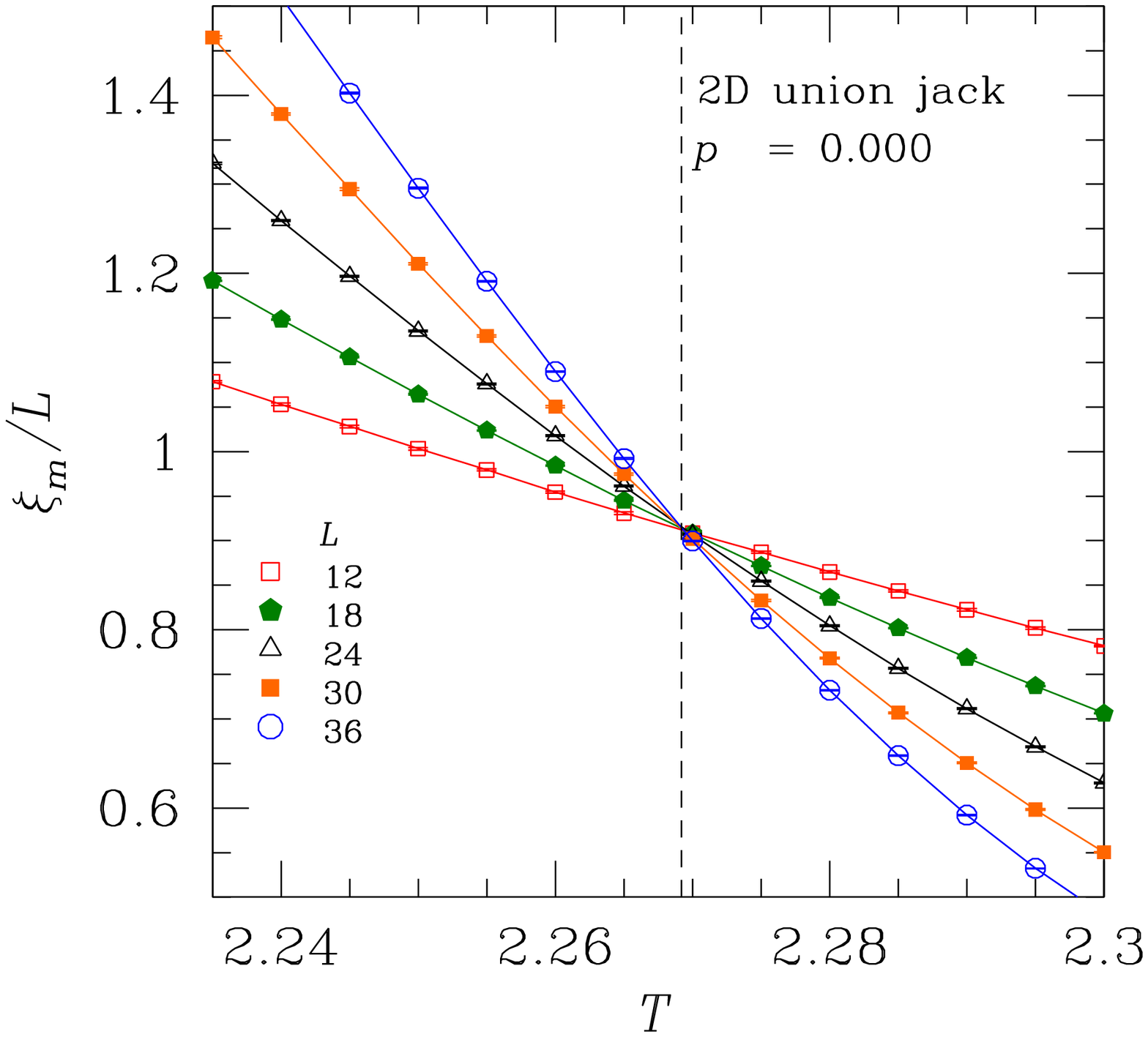}
\hspace*{-0.5cm}\includegraphics[width=6.0cm]{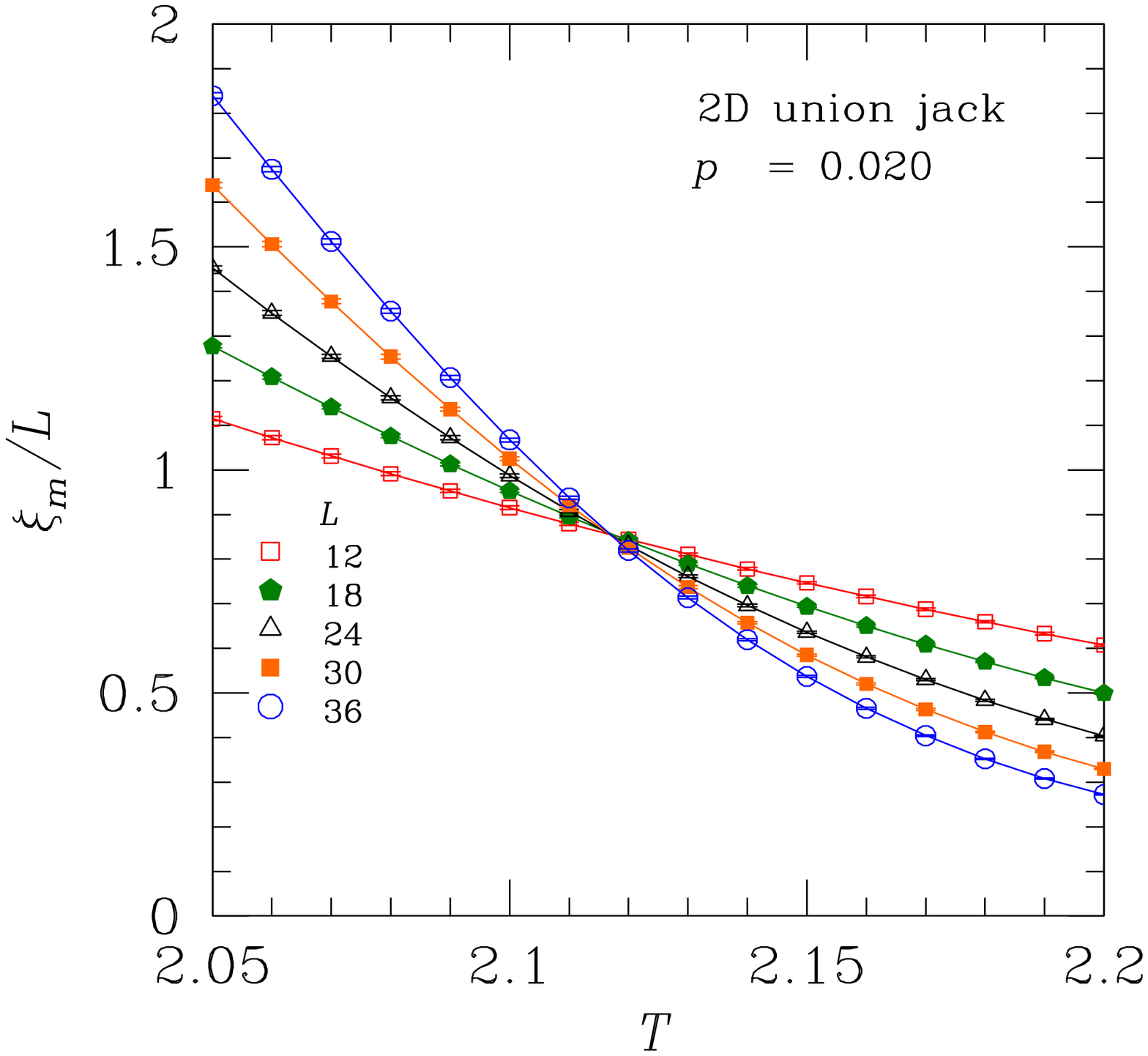}
\hspace*{-0.5cm}\includegraphics[width=6.0cm]{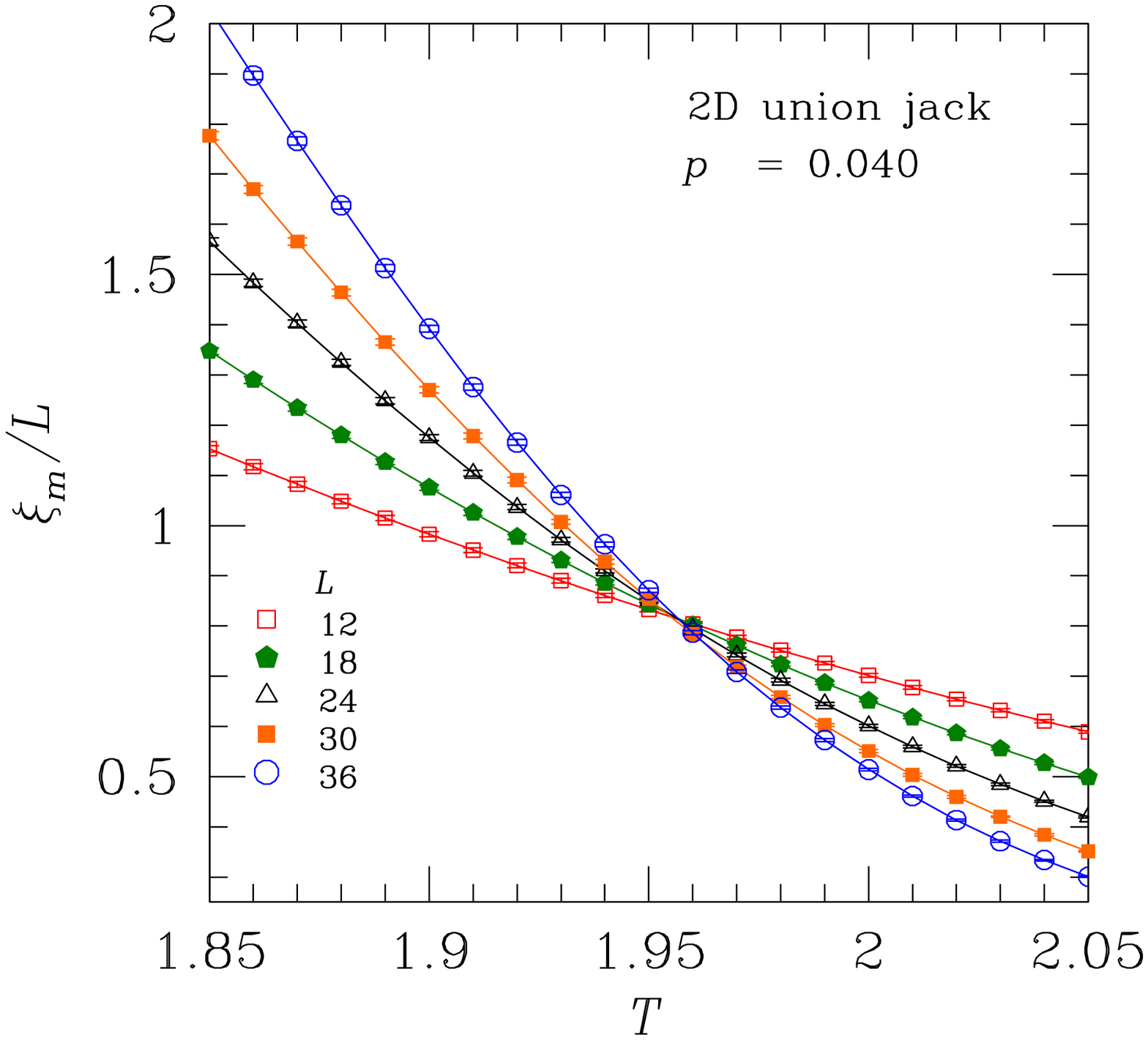}

\vspace*{-0.7cm}

\includegraphics[width=6.0cm]{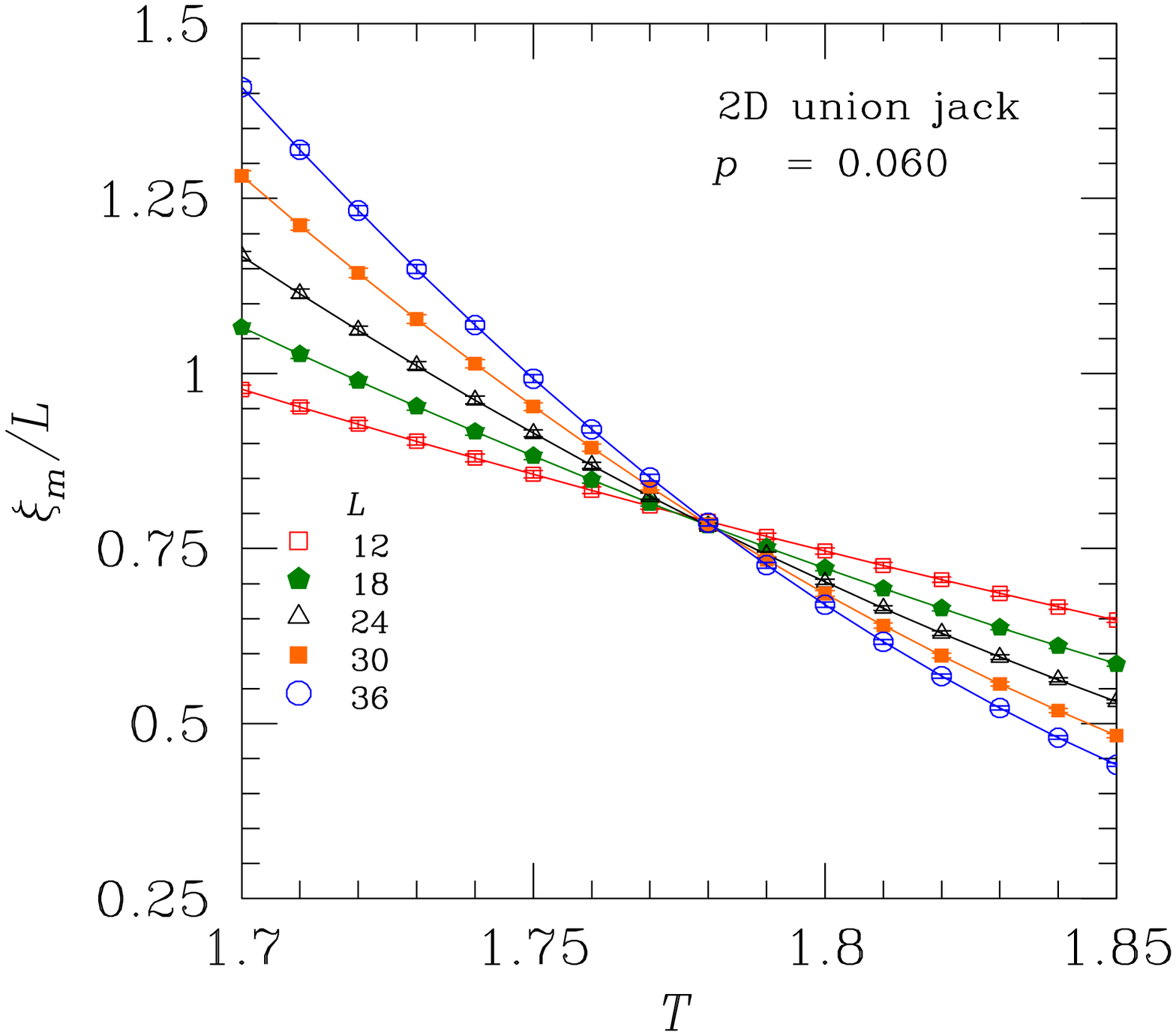}
\hspace*{-0.5cm}\includegraphics[width=6.0cm]{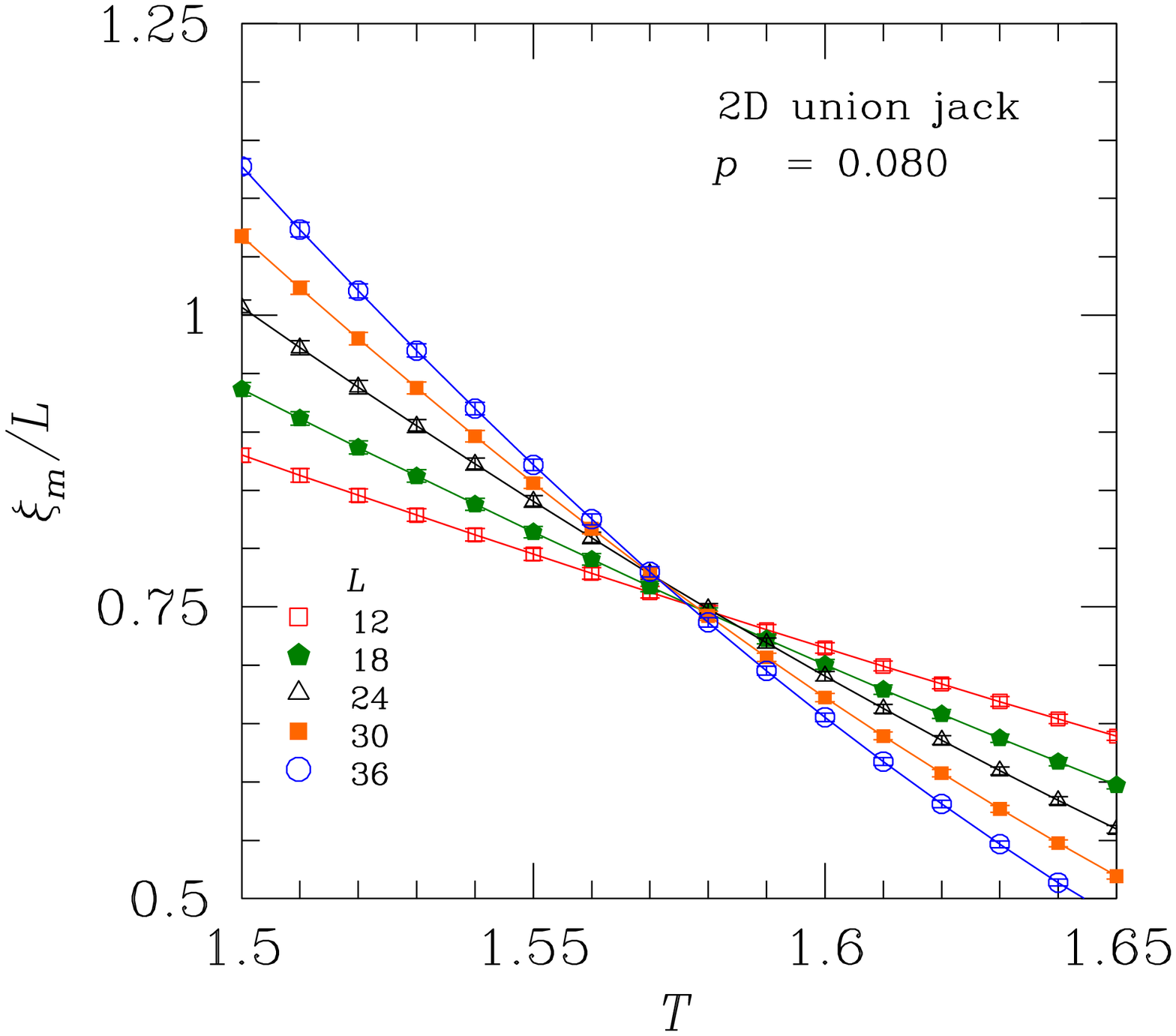}
\hspace*{-0.5cm}\includegraphics[width=6.0cm]{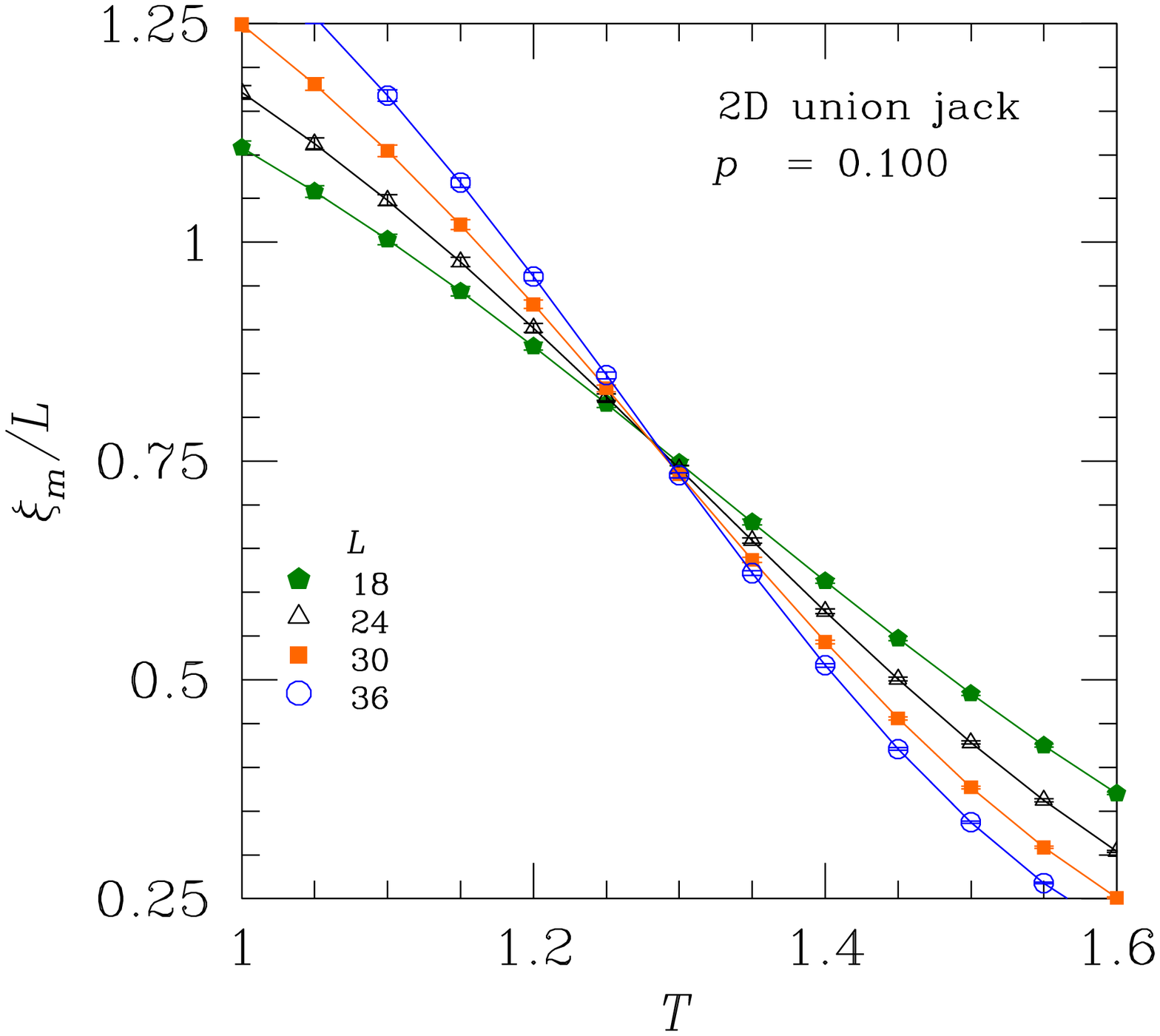}

\vspace*{-0.7cm}

\includegraphics[width=6.0cm]{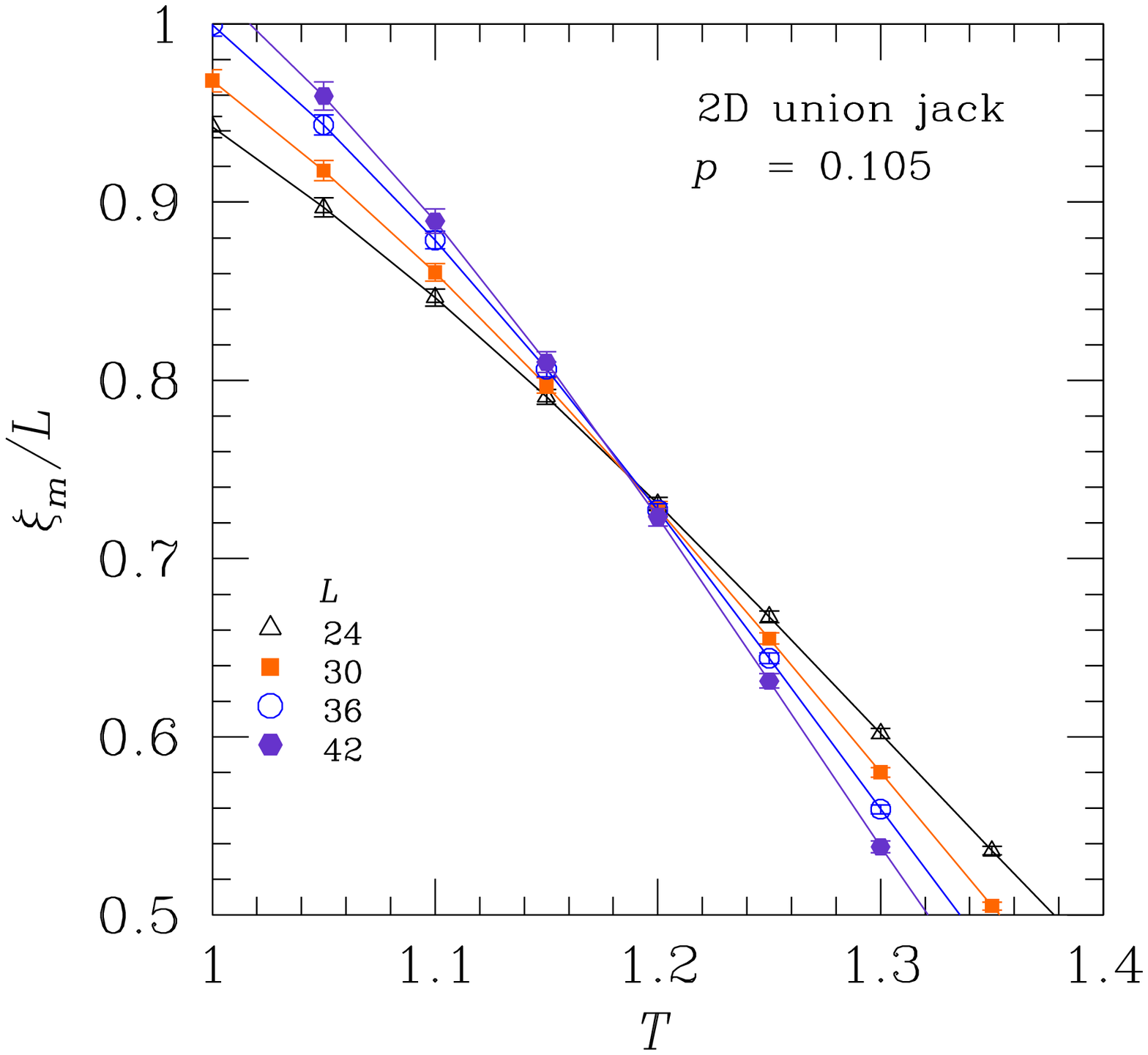}
\hspace*{-0.5cm}\includegraphics[width=6.0cm]{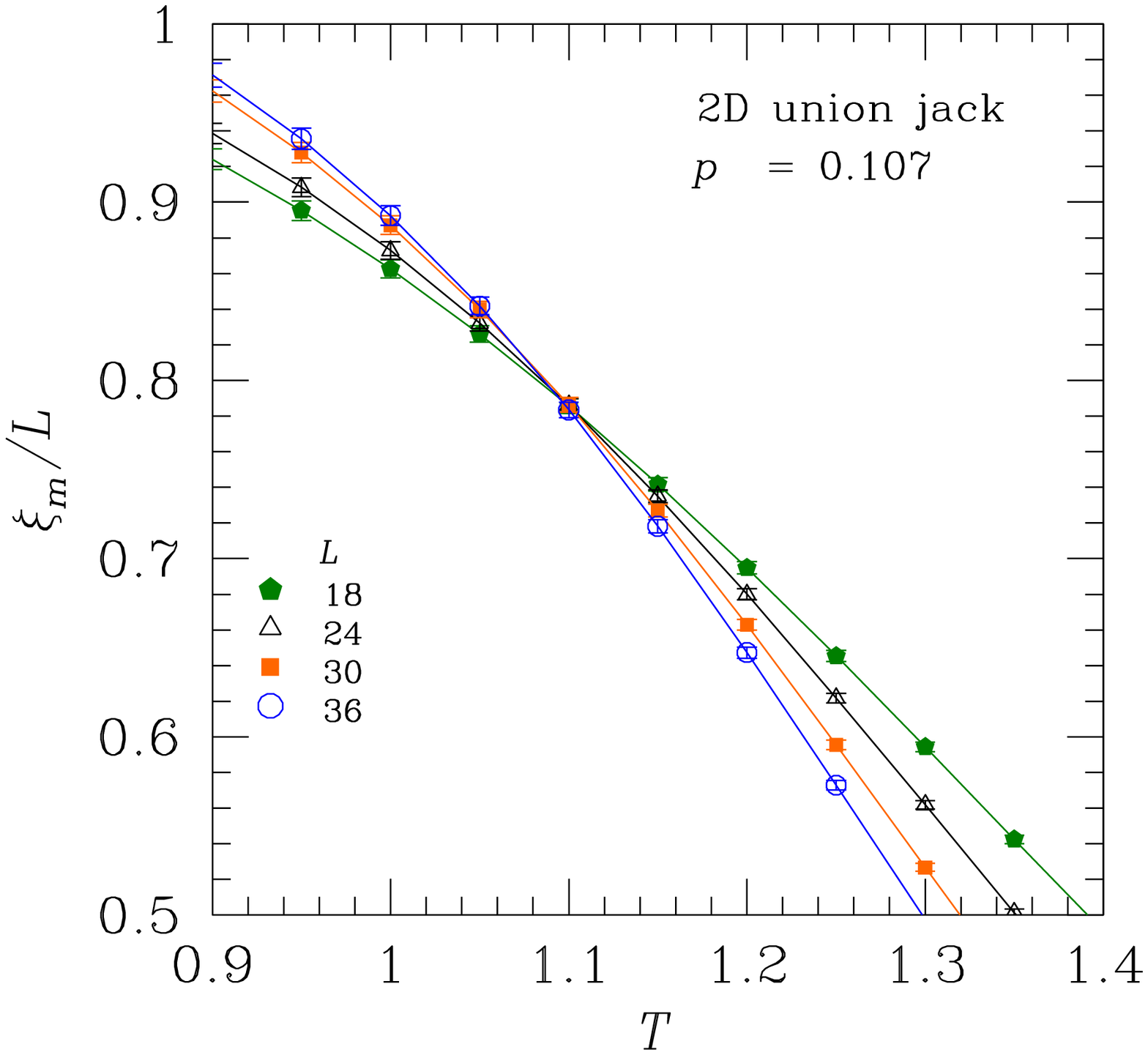}
\hspace*{-0.5cm}\includegraphics[width=6.0cm]{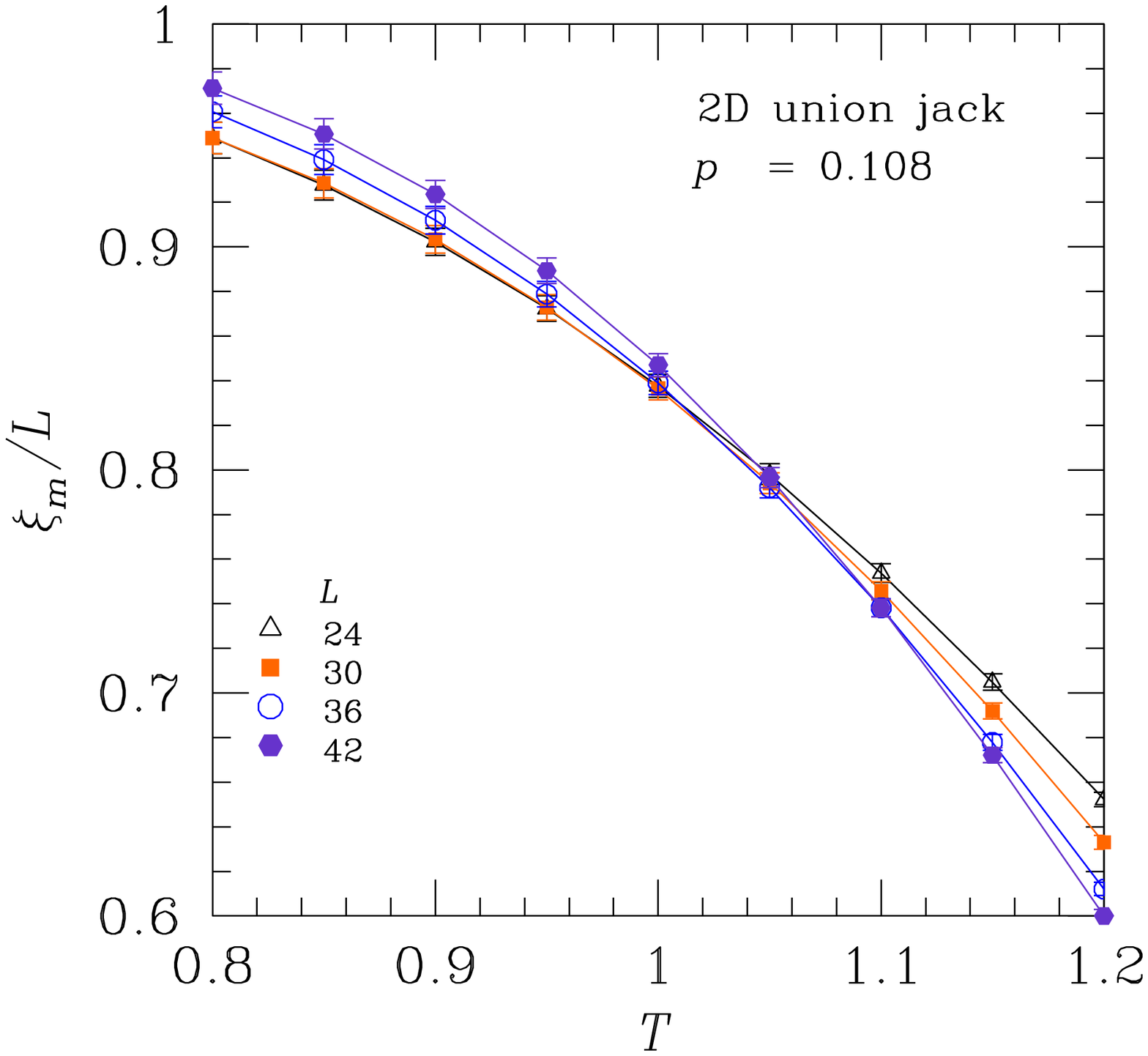}

\vspace*{-0.7cm}

\includegraphics[width=6.0cm]{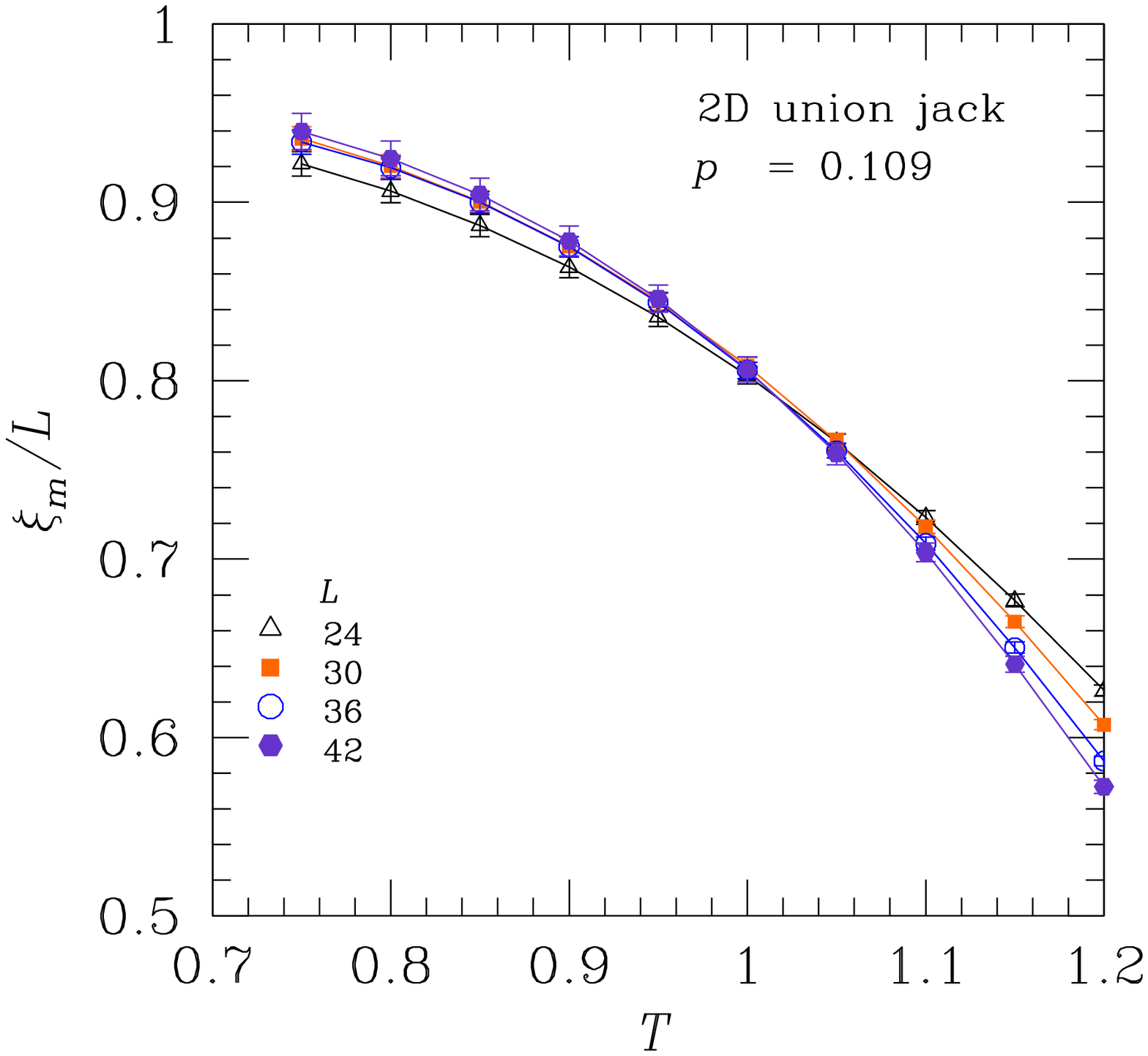}
\hspace*{-0.5cm}\includegraphics[width=6.0cm]{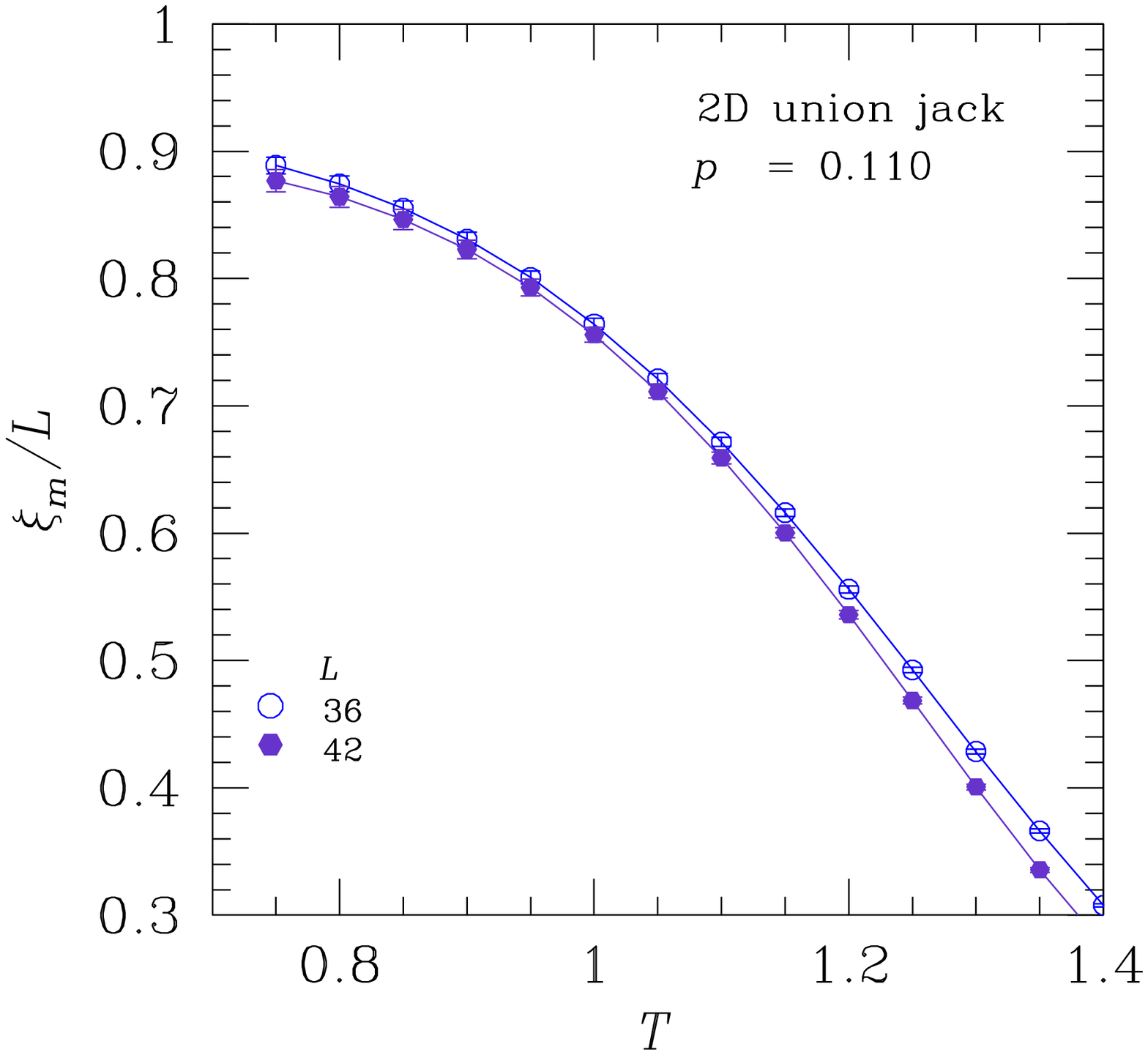}
\hspace*{-0.5cm}\includegraphics[width=6.0cm]{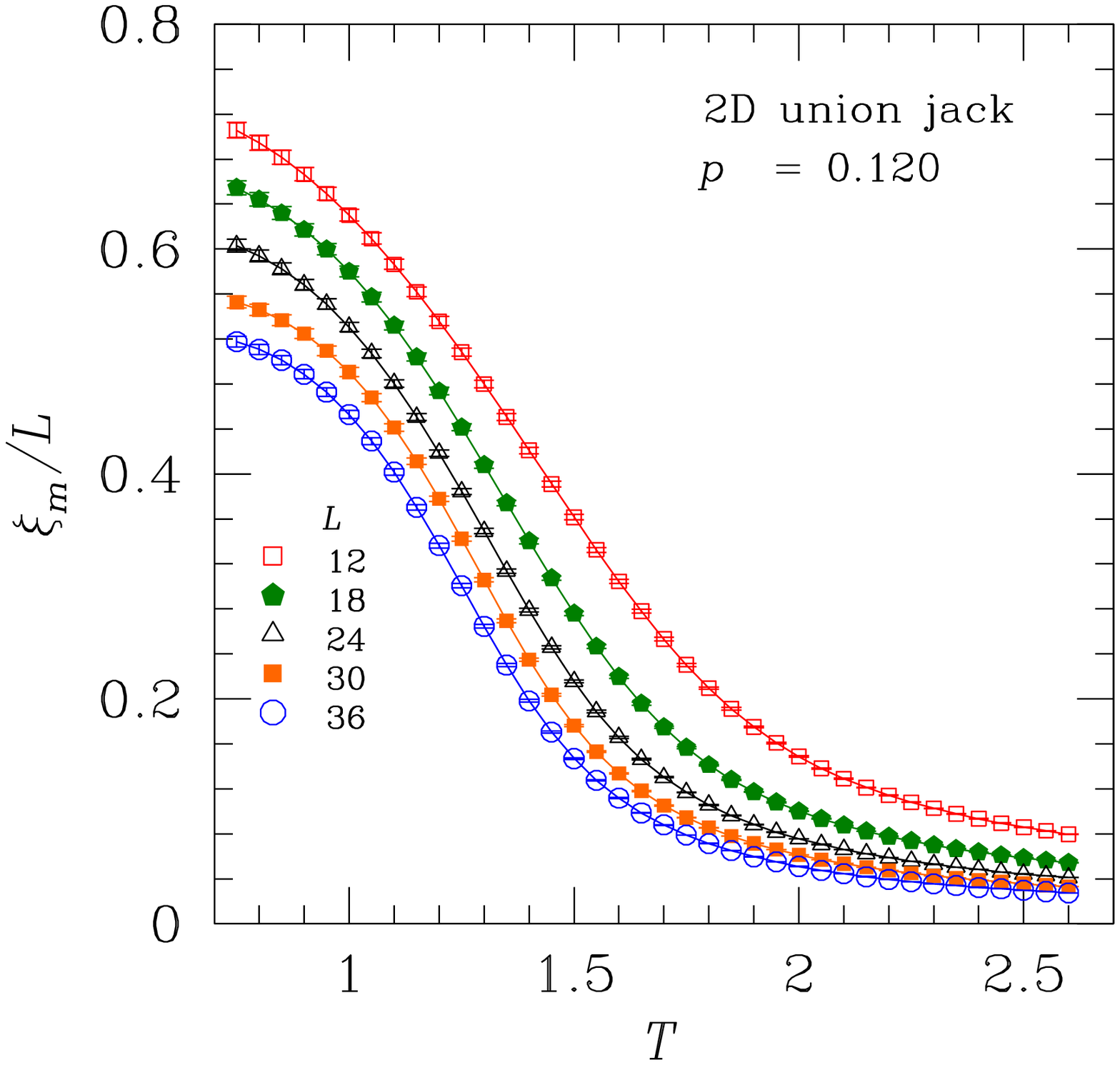}

\vspace*{-0.7cm}
\caption{(Color online)
Finite-size correlation length $\xi_{\rm m}/L$ as a function of
temperature $T$ for different values of $p$. The data for $p = 0$
cross at the critical temperature of the 2D Ising model (dashed line
top left panel). For $p \lesssim p_{\rm c} = 0.109$ there is signature
of a transition (data for different $L$ cross) whereas for $p  =
0.110 > p_{\rm c}$ the transition vanishes. Note that for $p \gtrsim 0.108$,
corrections are large thus making the determination of the transition
temperature difficult. For the case of $p = 0.110$ we only show the
two largest sizes for clarity.
}
\label{fig:cross}
\end{figure*}

The transitions to a ferromagnetic phase are determined by a
finite-size scaling of the dimensionless two-point finite-size
correlation length divided by the system size.  We start by determining
the wave-vector-dependent susceptibility
\begin{equation}
\chi(k) =  \frac{1}{L^2} \sum_{ij}^N \langle S_iS_j \rangle_T \;
           e^{i{\bf k}\cdot({\bf R}_i - {\bf R}_j)} \; .
\label{eq:chik}
\end{equation}
In Eq.~(\ref{eq:chik}) $\langle \cdots \rangle_T$ represents a
thermal average and ${\bf R}_i$ the spatial location of the spins.
The correlation length is then given by
\begin{equation}
\xi_{\rm m} = \frac{1}{2 \sin(k_{\rm min}/2)}
\sqrt{\frac{[\chi(k = 0)]_{\rm av}}{[\chi(k_{\rm min})]_{\rm av}}
- 1} ,
\label{eq:xiL}
\end{equation}
where $k_{\rm min} = (2 \pi / L,0)$ is the smallest nonzero wave vector
and $[\cdots]_{\rm av}$ represents an average over the different error
configurations (disorder sampling). The finite-size correlation length
divided by the system size has the following finite-size scaling form:
\begin{equation}
\xi_{\rm m}/L \sim \widetilde{X}(L^{1/\nu}[T - T_{\rm c}]) ,
\label{eq:fss}
\end{equation}
where $\nu$ is a critical exponent and $T_{\rm c}$ represents the
transition temperature. Numerically, finite systems of linear size
$L$ are studied. In that case the function $\xi_{\rm m}/L$ is
independent of $L$ whenever $T=T_{\rm c}$ as then the argument of the
function $\tilde{X}$ is zero.  In other words, when different system
sizes are studied, the data cross at one point, which corresponds to
the transition, if present.  This can be seen in Fig.\ref{fig:cross}.
Because finite-size scaling corrections are small in this case, one
can use the estimate of $T_{\rm c}$ obtained as a very good approximation
to the thermodynamic limit value.  The critical exponent $\nu$ for
the correlation length can be determined by a full scaling of the
data \cite{comment:scale}, as shown in Ref.~\cite{katzgraber:09c}.

We have also computed the spin-glass finite-size correlation length
[replace $S_i$ with $S_i^\alpha S_i^\beta$ in Eq.~(\ref{eq:chik})].
For all values of $p$ studied, no sign of a finite-temperature
spin-glass transition was found. This result coincides with the
results found for the triangular lattice \cite{katzgraber:09c}
and is compatible with the standard belief \cite{nishimori:01}
that no glassy phase is expected in two-dimensional random models,
although here this is extended to systems with three-body interactions.
We believe that a spin-glass phase may be possible for spin models
with random many-body interactions in three space dimensions.

Because the complexity of the problem increases considerable when $p >
0$, we use parallel tempering Monte Carlo \cite{geyer:91,hukushima:96}.
In addition to local spin flips, after each lattice sweep a global
update that exchanges the temperature of two replicas (copies) of
the system is proposed. This considerably speeds up the simulations
(parameters are shown in Table \ref{tab:simparams}). Equilibration
is tested by a logarithmic binning of the data. Once {\em at least}
the last three bins agree within errors, we define the system to be
equilibrated. Note that we use periodic boundary conditions to reduce
finite-size effects. To prevent a mismatch with the vertex coloring
rules of the lattice and boundary conditions, we use system sizes $L$
that are a multiple of $6$.

\begin{figure}

\includegraphics[width=\columnwidth]{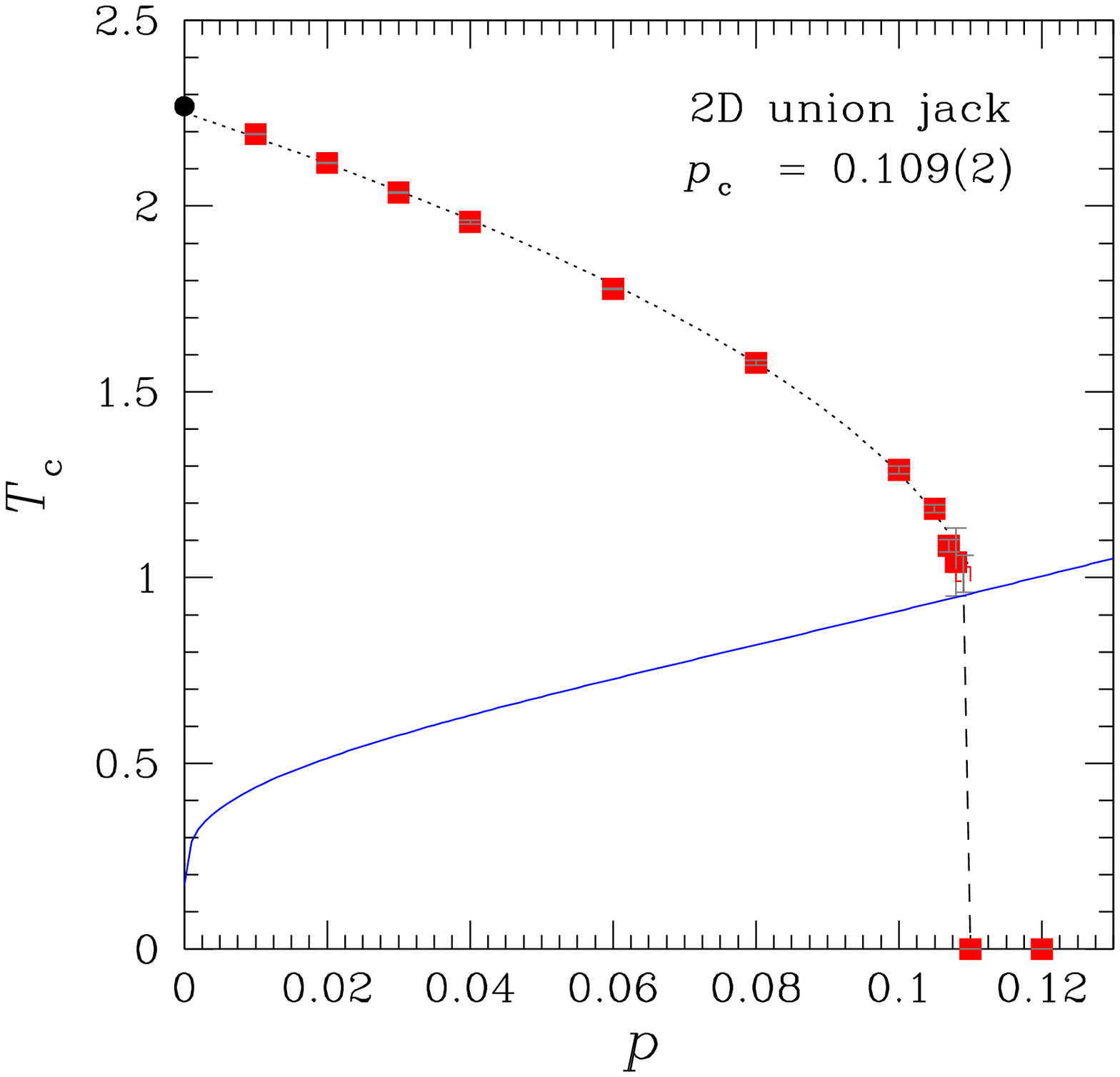}

\vspace*{-1.0cm}
\caption{(Color online)
$p\,$--$T_{\rm c}$ phase diagram for the random three-body Ising model on the
Union Jack lattice.  For $p > p_{\rm c} \approx 0.109$ the ferromagnetic
order is lost. The dotted line is a guide to the eye, the black
circle represents the analytically known transition temperature of
the 2D Ising model. The blue (solid) line represents the Nishimori
line. In the regime marked by a dashed line the exact determination
of $T_{\rm c}(p)$ is difficult.
}
\label{fig:pd}
\end{figure}

\section{Results}
\label{sec:results}

Figure \ref{fig:cross} shows the temperature-dependent dimensionless
finite-size correlation length for different values of $p$. The
top-left panel shows data for $p = 0$. The vertical dashed line is
the analytically known transition temperature of the two-dimensional
Ising model (i.e., $T_{\rm c} \simeq 2.2692$ \cite{yeomans:92}). It
is remarkable that the three-body Ising model on the Union Jack
lattice has the same transition temperature, albeit being in a
different universality class.  Its critical exponents are fully
determined by knowing two of them (e.g., $\nu = 3/4$ and $\alpha =
1/2$ \cite{comment:ising}). For $p \approx 0.108$ critical behavior
sets in, and for $p = 0.110$ no sign of a transition is visible.
We thus estimate $p_{\rm c} = 0.109(2)$, in agreement with results for
the triangular (TR) lattice \cite{katzgraber:09c} and the toric code
\cite{honecker:01,merz:02,ohzeki:09,parisen:09}.

\begin{figure}

\includegraphics[width=\columnwidth]{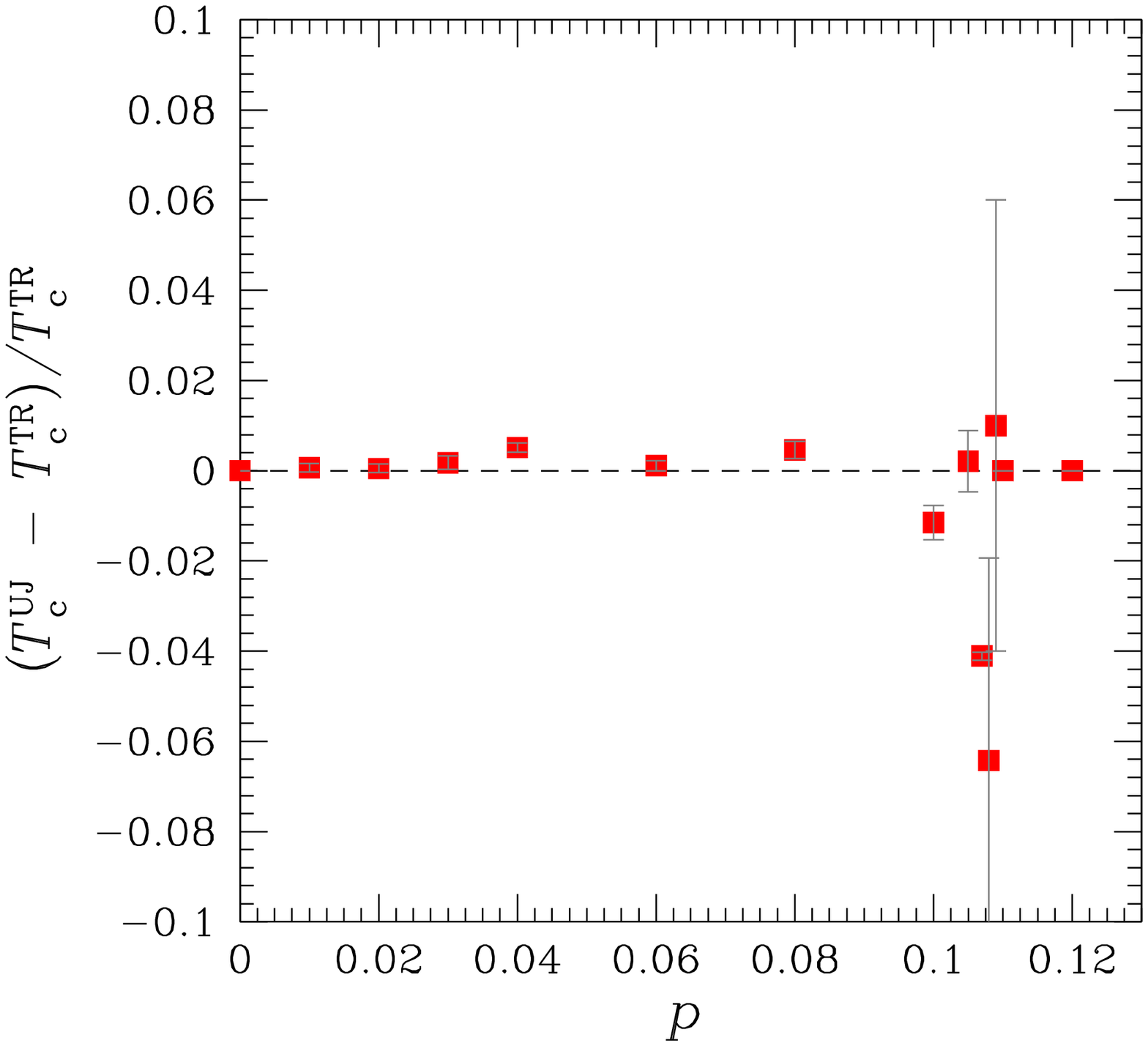}

\vspace*{-1.0cm}
\caption{(Color online)
Relative difference between $T_{\rm c}$ for the Union Jack (UJ) and
triangular lattices (TR) \cite{katzgraber:09c} as a function of $p$. In
most cases the deviations are smaller than 1\%. Around $p_{\rm c} \approx
0.109$ strong fluctuations appear since it is hard to estimate the
transition temperatures. The data for the triangular lattice have
been adapted from Ref.~\cite{katzgraber:09c} and expanded. Note that
for $p \to 0$ the deviations are smallest, suggesting that the slopes
of the phase lines for $p = 0$ are very close.
}
\label{fig:diff}
\end{figure}

The full $p\,$--$T_{\rm c}$ phase diagram for the UJ lattice is shown
in Fig.~\ref{fig:pd}; the solid (blue) line is the Nishimori line.
To check the differences between the phase diagram for the UJ and
TR lattices, we plot the relative deviation between the different
estimates of the critical temperature, $(T_{\rm c}^{\rm UJ} - T_{\rm c}^{\rm
TR})/T_{\rm c}^{\rm TR}$, as a function of $p$ (Fig.~\ref{fig:diff}).
For most values of $p$ the fluctuations are statistical and not larger
than 1\%. For $p \sim p_{\rm c}$, fluctuations are larger and the deviations
are of the order of $\sim 7$\%. Therefore, we believe that the phase
boundaries for both models are very close.

\section{Conclusions}
\label{sec:conlcusions}

In this work we have computed the $p\,$--$T$ phase diagram for the
random three-body Ising model on a two-dimensional Union Jack lattice
(Figs.~\ref{fig:lattice} and \ref{fig:pd}). The original motivation
for this numerical study with Monte Carlo methods is to compute the
multicritical point $p_{\rm c}$ in this phase diagram: the crossing point
between the critical line separating ordered/disordered phases and
the Nishimori line. The crossing point $p_{\rm c}$ corresponds to the error
threshold for topological color codes defined on the square-octagonal
lattice (the dual of which is the Union Jack lattice)---a value
that decides whether a topological stabilizer code is good enough
for performing quantum error correction in practical applications.
Our numerical result of $p_{\rm c}=0.109(2)$ is in agreement with the
corresponding value for the TCCs on triangular lattices and the
toric code, within error bars. This result shows that TCCs on the
square-octagonal lattice, which allow for the implementation of
the complete Clifford group of quantum gates \cite{bombin:06}, are
similarly stable as the toric code. The fact that the triangular
lattice and the UJ lattice share similar values of $p_{\rm c}$ that agree
within error bars is by no means obvious since both models are known
to be in different universality classes.  Note that the Monte Carlo
method combined with finite-size scaling used to determine the error
threshold does not rely on approximations. The computed values of
the phase boundary can merely be affected by finite system sizes,
although corrections to scaling are small. Furthermore, effects of
realistic error models have yet to be tested.

\begin{acknowledgments}

We would like to thank A.~Landahl and M.~Ohzeki for useful
discussions.  M.A.M.-D.~and H.B.~acknowledge financial support from
a PFI grant of EJ-GV, DGS grants under contracts, FIS2006-04885, and
the ESF INSTANS 2005-10.  H.G.K.~acknowledges support from the SNF
(Grant No.~PP002-114713).  The authors acknowledge Texas A\&M
University for access to their hydra cluster, the Texas Advanced
Computing Center (TACC) at The University of Texas at Austin for
providing HPC resources (Ranger Sun Constellation Linux Cluster), the
Centro de Supercomputaci{\'o}n y Visualizaci{\'o}n de Madrid (CeSViMa)
for access to the magerit cluster, the Barcelona Supercomputing Center
for access to the MareNostrum cluster within the Spanish Supercomputing
Network and ETH Zurich for CPU time on the Brutus cluster.

\end{acknowledgments}

\bibliography{refs,comments}

\begin{thebibliography}{46}
\expandafter\ifx\csname natexlab\endcsname\relax\def\natexlab#1{#1}\fi
\expandafter\ifx\csname bibnamefont\endcsname\relax
  \def\bibnamefont#1{#1}\fi
\expandafter\ifx\csname bibfnamefont\endcsname\relax
  \def\bibfnamefont#1{#1}\fi
\expandafter\ifx\csname citenamefont\endcsname\relax
  \def\citenamefont#1{#1}\fi
\expandafter\ifx\csname url\endcsname\relax
  \def\url#1{\texttt{#1}}\fi
\expandafter\ifx\csname urlprefix\endcsname\relax\def\urlprefix{URL }\fi
\providecommand{\bibinfo}[2]{#2}
\providecommand{\eprint}[2][]{\url{#2}}

\bibitem[{\citenamefont{Katzgraber et~al.}(2009)\citenamefont{Katzgraber,
  Bombin, and Martin-Delgado}}]{katzgraber:09c}
\bibinfo{author}{\bibfnamefont{H.~G.} \bibnamefont{Katzgraber}},
  \bibinfo{author}{\bibfnamefont{H.}~\bibnamefont{Bombin}}, \bibnamefont{and}
  \bibinfo{author}{\bibfnamefont{M.~A.} \bibnamefont{Martin-Delgado}},
  \emph{\bibinfo{title}{{Error Threshold for Color Codes and Random 3-Body
  Ising Models}}}, \bibinfo{journal}{Phys. Rev. Lett.}
  \textbf{\bibinfo{volume}{103}}, \bibinfo{pages}{090501}
  (\bibinfo{year}{2009}).

\bibitem[{\citenamefont{Ohzeki}(2009{\natexlab{a}})}]{ohzeki:09a}
\bibinfo{author}{\bibfnamefont{M.}~\bibnamefont{Ohzeki}},
  \emph{\bibinfo{title}{{Accuracy thresholds of topological color codes on the
  hexagonal and square-octagonal lattices}}}, \bibinfo{journal}{Phys. Rev. E}
  \textbf{\bibinfo{volume}{80}}, \bibinfo{pages}{011141}
  (\bibinfo{year}{2009}{\natexlab{a}}).

\bibitem[{\citenamefont{Kitaev}(2003)}]{kitaev:03}
\bibinfo{author}{\bibfnamefont{A.~Y.} \bibnamefont{Kitaev}},
  \emph{\bibinfo{title}{{Fault-tolerant quantum computation by anyons}}},
  \bibinfo{journal}{Ann. Phys.} \textbf{\bibinfo{volume}{303}},
  \bibinfo{pages}{2} (\bibinfo{year}{2003}).

\bibitem[{\citenamefont{Bombin and Martin-Delgado}(2006)}]{bombin:06}
\bibinfo{author}{\bibfnamefont{H.}~\bibnamefont{Bombin}} \bibnamefont{and}
  \bibinfo{author}{\bibfnamefont{M.~A.} \bibnamefont{Martin-Delgado}},
  \emph{\bibinfo{title}{{{Topological Quantum Distilation}}}},
  \bibinfo{journal}{Phys. Rev. Lett.} \textbf{\bibinfo{volume}{97}},
  \bibinfo{pages}{180501} (\bibinfo{year}{2006}).

\bibitem[{\citenamefont{Bombin and
  Martin-Delgado}(2007{\natexlab{a}})}]{bombin:07b}
\bibinfo{author}{\bibfnamefont{H.}~\bibnamefont{Bombin}} \bibnamefont{and}
  \bibinfo{author}{\bibfnamefont{M.~A.} \bibnamefont{Martin-Delgado}},
  \emph{\bibinfo{title}{{{Topological Computation without Braiding}}}},
  \bibinfo{journal}{Phys. Rev. Lett.} \textbf{\bibinfo{volume}{98}},
  \bibinfo{pages}{160502} (\bibinfo{year}{2007}{\natexlab{a}}).

\bibitem[{\citenamefont{Nayak et~al.}(2008)\citenamefont{Nayak, Simon, Stern,
  Freedman, and Das~Sarma}}]{nayak:08}
\bibinfo{author}{\bibfnamefont{C.}~\bibnamefont{Nayak}},
  \bibinfo{author}{\bibfnamefont{S.~H.} \bibnamefont{Simon}},
  \bibinfo{author}{\bibfnamefont{A.}~\bibnamefont{Stern}},
  \bibinfo{author}{\bibfnamefont{M.}~\bibnamefont{Freedman}}, \bibnamefont{and}
  \bibinfo{author}{\bibfnamefont{S.}~\bibnamefont{Das~Sarma}},
  \emph{\bibinfo{title}{{{Non-Abelian anyons and topological quantum
  computation}}}}, \bibinfo{journal}{Rev. Mod. Phys.}
  \textbf{\bibinfo{volume}{80}}, \bibinfo{pages}{1083} (\bibinfo{year}{2008}).

\bibitem[{\citenamefont{Bombin and
  Martin-Delgado}(2007{\natexlab{b}})}]{bombin:07c}
\bibinfo{author}{\bibfnamefont{H.}~\bibnamefont{Bombin}} \bibnamefont{and}
  \bibinfo{author}{\bibfnamefont{M.~A.} \bibnamefont{Martin-Delgado}},
  \emph{\bibinfo{title}{{{Optimal resources for topological two-dimensional
  stabilizer codes: Comparative study}}}}, \bibinfo{journal}{Phys. Rev. A}
  \textbf{\bibinfo{volume}{76}}, \bibinfo{pages}{012305}
  (\bibinfo{year}{2007}{\natexlab{b}}).

\bibitem[{\citenamefont{Honecker et~al.}(2001)\citenamefont{Honecker, Picco,
  and Pujol}}]{honecker:01}
\bibinfo{author}{\bibfnamefont{A.}~\bibnamefont{Honecker}},
  \bibinfo{author}{\bibfnamefont{M.}~\bibnamefont{Picco}}, \bibnamefont{and}
  \bibinfo{author}{\bibfnamefont{P.}~\bibnamefont{Pujol}},
  \emph{\bibinfo{title}{{Universality Class of the Nishimori Point in the 2D
  {$\pm J$} Random-Bond Ising Model}}}, \bibinfo{journal}{Phys. Rev. Lett.}
  \textbf{\bibinfo{volume}{87}}, \bibinfo{pages}{047201}
  (\bibinfo{year}{2001}).

\bibitem[{\citenamefont{Merz and Chalker}(2002)}]{merz:02}
\bibinfo{author}{\bibfnamefont{F.}~\bibnamefont{Merz}} \bibnamefont{and}
  \bibinfo{author}{\bibfnamefont{J.~T.} \bibnamefont{Chalker}},
  \emph{\bibinfo{title}{{Two-dimensional random-bond Ising model, free
  fermions, and the network model}}}, \bibinfo{journal}{Phys. Rev. B}
  \textbf{\bibinfo{volume}{65}}, \bibinfo{pages}{054425}
  (\bibinfo{year}{2002}).

\bibitem[{\citenamefont{Ohzeki}(2009{\natexlab{b}})}]{ohzeki:09}
\bibinfo{author}{\bibfnamefont{M.}~\bibnamefont{Ohzeki}},
  \emph{\bibinfo{title}{{{Precise locations of multicritical points for spin
  glasses on regular lattices}}}}, \bibinfo{journal}{Phys. Rev. E}
  \textbf{\bibinfo{volume}{79}}, \bibinfo{pages}{021129}
  (\bibinfo{year}{2009}{\natexlab{b}}).

\bibitem[{\citenamefont{{Parisen Toldin} et~al.}(2009)\citenamefont{{Parisen
  Toldin}, {Pelissetto}, and {Vicari}}}]{parisen:09}
\bibinfo{author}{\bibfnamefont{F.}~\bibnamefont{{Parisen Toldin}}},
  \bibinfo{author}{\bibfnamefont{A.}~\bibnamefont{{Pelissetto}}},
  \bibnamefont{and} \bibinfo{author}{\bibfnamefont{E.}~\bibnamefont{{Vicari}}},
  \emph{\bibinfo{title}{{{Strong-Disorder Paramagnetic-Ferromagnetic Fixed
  Point in the Square-Lattice {$\pm J$} Ising Model}}}}, \bibinfo{journal}{J.
  Stat. Phys.} \textbf{\bibinfo{volume}{135}}, \bibinfo{pages}{1039}
  (\bibinfo{year}{2009}).

\bibitem[{\citenamefont{Wang et~al.}(2009)\citenamefont{Wang, Fowler, Hill, and
  Hollenberg}}]{wang:09}
\bibinfo{author}{\bibfnamefont{D.~S.} \bibnamefont{Wang}},
  \bibinfo{author}{\bibfnamefont{A.~G.} \bibnamefont{Fowler}},
  \bibinfo{author}{\bibfnamefont{C.~D.} \bibnamefont{Hill}}, \bibnamefont{and}
  \bibinfo{author}{\bibfnamefont{L.~C.~L.} \bibnamefont{Hollenberg}},
  \emph{\bibinfo{title}{{{Graphical algorithms and threshold error rates for
  the 2d colour code}}}} (\bibinfo{year}{2009}),
  \bibinfo{note}{(arXiv:0907.1708)}.

\bibitem[{\citenamefont{Landahl et~al.}()\citenamefont{Landahl, Anderson, and
  Rice}}]{landahl:09}
\bibinfo{author}{\bibfnamefont{A.}~\bibnamefont{Landahl}},
  \bibinfo{author}{\bibfnamefont{J.~T.} \bibnamefont{Anderson}},
  \bibnamefont{and} \bibinfo{author}{\bibfnamefont{P.}~\bibnamefont{Rice}},
  \bibinfo{note}{in preparation (2009)}.

\bibitem[{\citenamefont{Dennis et~al.}(2002)\citenamefont{Dennis, Kitaev,
  Landahl, and Preskill}}]{dennis:02}
\bibinfo{author}{\bibfnamefont{E.}~\bibnamefont{Dennis}},
  \bibinfo{author}{\bibfnamefont{A.}~\bibnamefont{Kitaev}},
  \bibinfo{author}{\bibfnamefont{A.}~\bibnamefont{Landahl}}, \bibnamefont{and}
  \bibinfo{author}{\bibfnamefont{J.}~\bibnamefont{Preskill}},
  \emph{\bibinfo{title}{{Topological quantum memory}}}, \bibinfo{journal}{J.
  Math. Phys.} \textbf{\bibinfo{volume}{43}}, \bibinfo{pages}{4452}
  (\bibinfo{year}{2002}).

\bibitem[{\citenamefont{Bombin and
  Martin-Delgado}(2007{\natexlab{c}})}]{bombin:07}
\bibinfo{author}{\bibfnamefont{H.}~\bibnamefont{Bombin}} \bibnamefont{and}
  \bibinfo{author}{\bibfnamefont{M.~A.} \bibnamefont{Martin-Delgado}},
  \emph{\bibinfo{title}{{{Exact topological quantum order in $D=3$ and beyond:
  Branyons and brane-net condensates}}}}, \bibinfo{journal}{Phys. Rev. B}
  \textbf{\bibinfo{volume}{75}}, \bibinfo{pages}{075103}
  (\bibinfo{year}{2007}{\natexlab{c}}).

\bibitem[{\citenamefont{Bravyi and Raussendorf}(2007)}]{bravyi:07}
\bibinfo{author}{\bibfnamefont{S.}~\bibnamefont{Bravyi}} \bibnamefont{and}
  \bibinfo{author}{\bibfnamefont{R.}~\bibnamefont{Raussendorf}},
  \emph{\bibinfo{title}{{{Measurement-based quantum computation with the toric
  code states}}}}, \bibinfo{journal}{Phys. Rev. A}
  \textbf{\bibinfo{volume}{76}}, \bibinfo{pages}{022304}
  (\bibinfo{year}{2007}).

\bibitem[{\citenamefont{Bombin and Martin-Delgado}(2008)}]{bombin:08}
\bibinfo{author}{\bibfnamefont{H.}~\bibnamefont{Bombin}} \bibnamefont{and}
  \bibinfo{author}{\bibfnamefont{M.~A.} \bibnamefont{Martin-Delgado}},
  \emph{\bibinfo{title}{{{Statistical mechanical models and topological color
  codes}}}}, \bibinfo{journal}{Phys. Rev. A} \textbf{\bibinfo{volume}{77}},
  \bibinfo{pages}{042322} (\bibinfo{year}{2008}).

\bibitem[{\citenamefont{Van~den Nest et~al.}(2008)\citenamefont{Van~den Nest,
  D{\"u}r, and Briegel}}]{vandennest:08}
\bibinfo{author}{\bibfnamefont{M.}~\bibnamefont{Van~den Nest}},
  \bibinfo{author}{\bibfnamefont{W.}~\bibnamefont{D{\"u}r}}, \bibnamefont{and}
  \bibinfo{author}{\bibfnamefont{H.~J.} \bibnamefont{Briegel}},
  \emph{\bibinfo{title}{{{Completeness of the Classical 2D Ising Model and
  Universal Quantum Computation}}}}, \bibinfo{journal}{Phys. Rev. Lett.}
  \textbf{\bibinfo{volume}{100}}, \bibinfo{pages}{110501}
  (\bibinfo{year}{2008}).

\bibitem[{\citenamefont{{de Las Cuevas} et~al.}(2009)\citenamefont{{de Las
  Cuevas}, D{\"u}r, Briegel, and Martin-Delgado}}]{delascuevas:09}
\bibinfo{author}{\bibfnamefont{G.}~\bibnamefont{{de Las Cuevas}}},
  \bibinfo{author}{\bibfnamefont{W.}~\bibnamefont{D{\"u}r}},
  \bibinfo{author}{\bibfnamefont{H.~J.} \bibnamefont{Briegel}},
  \bibnamefont{and} \bibinfo{author}{\bibfnamefont{M.~A.}
  \bibnamefont{Martin-Delgado}}, \emph{\bibinfo{title}{{{Unifying All Classical
  Spin Models in a Lattice Gauge Theory}}}}, \bibinfo{journal}{Phys. Rev.
  Lett.} \textbf{\bibinfo{volume}{102}}, \bibinfo{pages}{230502}
  (\bibinfo{year}{2009}).

\bibitem[{\citenamefont{Yeomans}(1992)}]{yeomans:92}
\bibinfo{author}{\bibfnamefont{J.~M.} \bibnamefont{Yeomans}},
  \emph{\bibinfo{title}{{Statistical Mechanics of Phase Transitions}}}
  (\bibinfo{publisher}{Oxford University Press}, \bibinfo{address}{Oxford},
  \bibinfo{year}{1992}).

\bibitem[{\citenamefont{Cardy}(1996)}]{cardy:96}
\bibinfo{author}{\bibfnamefont{J.}~\bibnamefont{Cardy}},
  \emph{\bibinfo{title}{{Scaling and Renormalization in Statistical Physics}}}
  (\bibinfo{publisher}{Cambridge University Press},
  \bibinfo{address}{Cambridge}, \bibinfo{year}{1996}).

\bibitem[{\citenamefont{Hintermann and Merlini}(1972)}]{hintermann:72}
\bibinfo{author}{\bibfnamefont{A.}~\bibnamefont{Hintermann}} \bibnamefont{and}
  \bibinfo{author}{\bibfnamefont{D.}~\bibnamefont{Merlini}},
  \emph{\bibinfo{title}{{{Exact solution of a two dimensional Ising model with
  pure 3 spin interactions}}}}, \bibinfo{journal}{Phys. Lett. A}
  \textbf{\bibinfo{volume}{41}}, \bibinfo{pages}{208} (\bibinfo{year}{1972}).

\bibitem[{\citenamefont{{Baxter} and {Wu}}(1973)}]{baxter:73}
\bibinfo{author}{\bibfnamefont{R.~J.} \bibnamefont{{Baxter}}} \bibnamefont{and}
  \bibinfo{author}{\bibfnamefont{F.~Y.} \bibnamefont{{Wu}}},
  \emph{\bibinfo{title}{{{Exact Solution of an Ising Model with Three-Spin
  Interactions on a Triangular Lattice}}}}, \bibinfo{journal}{Phys. Rev. Lett.}
  \textbf{\bibinfo{volume}{31}}, \bibinfo{pages}{1294} (\bibinfo{year}{1973}).

\bibitem[{\citenamefont{Shor}(1995)}]{shor:95}
\bibinfo{author}{\bibfnamefont{P.~W.} \bibnamefont{Shor}},
  \emph{\bibinfo{title}{{{Scheme for reducing decoherence in quantum computer
  memory}}}}, \bibinfo{journal}{Phys. Rev. A} \textbf{\bibinfo{volume}{52}},
  \bibinfo{pages}{R2493} (\bibinfo{year}{1995}).

\bibitem[{\citenamefont{Steane}(1996{\natexlab{a}})}]{steane:96}
\bibinfo{author}{\bibfnamefont{A.~M.} \bibnamefont{Steane}},
  \emph{\bibinfo{title}{{{Error Correcting Codes in Quantum Theory}}}},
  \bibinfo{journal}{Phys. Rev. Lett.} \textbf{\bibinfo{volume}{77}},
  \bibinfo{pages}{793} (\bibinfo{year}{1996}{\natexlab{a}}).

\bibitem[{\citenamefont{Calderbank and Shor}(1996)}]{calderbank:96}
\bibinfo{author}{\bibfnamefont{A.~R.} \bibnamefont{Calderbank}}
  \bibnamefont{and} \bibinfo{author}{\bibfnamefont{P.~W.} \bibnamefont{Shor}},
  \emph{\bibinfo{title}{{Good quantum error-correcting codes exist}}},
  \bibinfo{journal}{Phys. Rev. A} \textbf{\bibinfo{volume}{54}},
  \bibinfo{pages}{1098} (\bibinfo{year}{1996}).

\bibitem[{\citenamefont{Steane}(1996{\natexlab{b}})}]{steane:96a}
\bibinfo{author}{\bibfnamefont{A.}~\bibnamefont{Steane}},
  \emph{\bibinfo{title}{{{Multiple-Particle Interference and Quantum Error
  Correction}}}}, \bibinfo{journal}{Proc. Roy. Soc. Lond. A}
  \textbf{\bibinfo{volume}{452}}, \bibinfo{pages}{2551}
  (\bibinfo{year}{1996}{\natexlab{b}}).

\bibitem[{\citenamefont{Gottesman}(1996)}]{gottesman:96}
\bibinfo{author}{\bibfnamefont{D.}~\bibnamefont{Gottesman}},
  \emph{\bibinfo{title}{{{Class of quantum error-correcting codes saturating
  the quantum Hamming bound}}}}, \bibinfo{journal}{Phys. Rev. A}
  \textbf{\bibinfo{volume}{54}}, \bibinfo{pages}{1862} (\bibinfo{year}{1996}).

\bibitem[{\citenamefont{Bombin and
  Martin-Delgado}(2007{\natexlab{d}})}]{bombin:07d}
\bibinfo{author}{\bibfnamefont{H.}~\bibnamefont{Bombin}} \bibnamefont{and}
  \bibinfo{author}{\bibfnamefont{M.~A.} \bibnamefont{Martin-Delgado}},
  \emph{\bibinfo{title}{{{Homological error correction: Classical and quantum
  codes}}}}, \bibinfo{journal}{J. Math. Phys.} \textbf{\bibinfo{volume}{48}},
  \bibinfo{pages}{052105} (\bibinfo{year}{2007}{\natexlab{d}}).

\bibitem[{\citenamefont{{Gao} et~al.}(2009)\citenamefont{{Gao}, {Fowler},
  {Raussendorf}, {Yao}, {Lu}, {Xu}, {Lu}, {Peng}, {Deng}, {Chen}
  et~al.}}]{gao:09}
\bibinfo{author}{\bibfnamefont{W.-B.} \bibnamefont{{Gao}}},
  \bibinfo{author}{\bibfnamefont{A.~G.} \bibnamefont{{Fowler}}},
  \bibinfo{author}{\bibfnamefont{R.}~\bibnamefont{{Raussendorf}}},
  \bibinfo{author}{\bibfnamefont{X.-C.} \bibnamefont{{Yao}}},
  \bibinfo{author}{\bibfnamefont{H.}~\bibnamefont{{Lu}}},
  \bibinfo{author}{\bibfnamefont{P.}~\bibnamefont{{Xu}}},
  \bibinfo{author}{\bibfnamefont{C.-Y.} \bibnamefont{{Lu}}},
  \bibinfo{author}{\bibfnamefont{C.-Z.} \bibnamefont{{Peng}}},
  \bibinfo{author}{\bibfnamefont{Y.}~\bibnamefont{{Deng}}},
  \bibinfo{author}{\bibfnamefont{Z.-B.} \bibnamefont{{Chen}}},
  \bibnamefont{et~al.}, \emph{\bibinfo{title}{{{Experimental demonstration of
  topological error correction}}}} (\bibinfo{year}{2009}),
  \bibinfo{note}{(arXiv:0905.1542)}.

\bibitem[{\citenamefont{Weimer et~al.}(2009)\citenamefont{Weimer, M{\"u}ller,
  Lesanovsky, Zoller, and B{\"u}chler}}]{weimer:09}
\bibinfo{author}{\bibfnamefont{H.}~\bibnamefont{Weimer}},
  \bibinfo{author}{\bibfnamefont{M.}~\bibnamefont{M{\"u}ller}},
  \bibinfo{author}{\bibfnamefont{I.}~\bibnamefont{Lesanovsky}},
  \bibinfo{author}{\bibfnamefont{P.}~\bibnamefont{Zoller}}, \bibnamefont{and}
  \bibinfo{author}{\bibfnamefont{H.~P.} \bibnamefont{B{\"u}chler}},
  \emph{\bibinfo{title}{{{Digital Coherent and Dissipative Quantum Simulations
  with Rydberg Atoms}}}} (\bibinfo{year}{2009}),
  \bibinfo{note}{(arXiv:0907.1657)}.

\bibitem[{\citenamefont{Bombin et~al.}(2009)\citenamefont{Bombin, Chhajlany,
  Horodecki, and Martin-Delgado}}]{bombin:09a}
\bibinfo{author}{\bibfnamefont{H.}~\bibnamefont{Bombin}},
  \bibinfo{author}{\bibfnamefont{R.~W.} \bibnamefont{Chhajlany}},
  \bibinfo{author}{\bibfnamefont{M.}~\bibnamefont{Horodecki}},
  \bibnamefont{and} \bibinfo{author}{\bibfnamefont{M.~A.}
  \bibnamefont{Martin-Delgado}}, \emph{\bibinfo{title}{{{Self-Correcting
  Quantum Computers}}}} (\bibinfo{year}{2009}),
  \bibinfo{note}{(arXiv:0907.5228)}.

\bibitem[{\citenamefont{Alicki et~al.}(2008)\citenamefont{Alicki, Horodecki,
  Horodecki, and Horodecki}}]{alicki:08}
\bibinfo{author}{\bibfnamefont{R.}~\bibnamefont{Alicki}},
  \bibinfo{author}{\bibfnamefont{M.}~\bibnamefont{Horodecki}},
  \bibinfo{author}{\bibfnamefont{P.}~\bibnamefont{Horodecki}},
  \bibnamefont{and}
  \bibinfo{author}{\bibfnamefont{R.}~\bibnamefont{Horodecki}},
  \emph{\bibinfo{title}{{{On thermal stability of topological qubit in Kitaev's
  4D model}}}} (\bibinfo{year}{2008}),
  \bibinfo{note}{(arXiv:quant-phys/0811.0033)}.

\bibitem[{\citenamefont{Bravyi and Kitaev}(1998)}]{bravyi:98}
\bibinfo{author}{\bibfnamefont{S.}~\bibnamefont{Bravyi}} \bibnamefont{and}
  \bibinfo{author}{\bibfnamefont{A.~Y.} \bibnamefont{Kitaev}},
  \emph{\bibinfo{title}{{{Quantum codes on a lattice with boundary}}}}
  (\bibinfo{year}{1998}), \bibinfo{note}{(arXiv:quant-phys/9811052)}.

\bibitem[{\citenamefont{Bombin and Martin-Delgado}(2009)}]{bombin:09}
\bibinfo{author}{\bibfnamefont{H.}~\bibnamefont{Bombin}} \bibnamefont{and}
  \bibinfo{author}{\bibfnamefont{M.~A.} \bibnamefont{Martin-Delgado}},
  \emph{\bibinfo{title}{{{Quantum measurements and gates by code
  deformation}}}}, \bibinfo{journal}{J. Phys. A} \textbf{\bibinfo{volume}{42}},
  \bibinfo{pages}{095302} (\bibinfo{year}{2009}).

\bibitem[{\citenamefont{Bombin}(2009)}]{bombin:09b}
\bibinfo{author}{\bibfnamefont{H.}~\bibnamefont{Bombin}},
  \emph{\bibinfo{title}{{{Topological Subsystem Codes}}}}
  (\bibinfo{year}{2009}), \bibinfo{note}{(arXiv:0908.4246)}.

\bibitem[{com({\natexlab{a}})}]{comment:J}
\bibinfo{note}{Without loss of generality, we set the energy scale $J = 1$.
  Therefore, temperatures are dimensionless.}

\bibitem[{\citenamefont{{Nishimori}}(1981)}]{nishimori:81}
\bibinfo{author}{\bibfnamefont{H.}~\bibnamefont{{Nishimori}}},
  \emph{\bibinfo{title}{{{Internal Energy, Specific Heat and Correlation
  Function of the Bond-Random Ising Model}}}}, \bibinfo{journal}{Prog. Theor.
  Phys.} \textbf{\bibinfo{volume}{66}}, \bibinfo{pages}{1169}
  (\bibinfo{year}{1981}).

\bibitem[{\citenamefont{Nishimori}(2001)}]{nishimori:01}
\bibinfo{author}{\bibfnamefont{H.}~\bibnamefont{Nishimori}},
  \emph{\bibinfo{title}{{Statistical Physics of Spin Glasses and Information
  Processing: An Introduction}}} (\bibinfo{publisher}{Oxford University Press},
  \bibinfo{address}{New York}, \bibinfo{year}{2001}).

\bibitem[{\citenamefont{Binder and Young}(1986)}]{binder:86}
\bibinfo{author}{\bibfnamefont{K.}~\bibnamefont{Binder}} \bibnamefont{and}
  \bibinfo{author}{\bibfnamefont{A.~P.} \bibnamefont{Young}},
  \emph{\bibinfo{title}{Spin glasses: Experimental facts, theoretical concepts
  and open questions}}, \bibinfo{journal}{Rev. Mod. Phys.}
  \textbf{\bibinfo{volume}{58}}, \bibinfo{pages}{801} (\bibinfo{year}{1986}).

\bibitem[{\citenamefont{Palassini and Caracciolo}(1999)}]{palassini:99b}
\bibinfo{author}{\bibfnamefont{M.}~\bibnamefont{Palassini}} \bibnamefont{and}
  \bibinfo{author}{\bibfnamefont{S.}~\bibnamefont{Caracciolo}},
  \emph{\bibinfo{title}{{U}niversal {F}inite-{S}ize {S}caling {F}unctions in
  the 3{D} {I}sing {S}pin {G}lass}}, \bibinfo{journal}{Phys. Rev. Lett.}
  \textbf{\bibinfo{volume}{82}}, \bibinfo{pages}{5128} (\bibinfo{year}{1999}).

\bibitem[{com({\natexlab{b}})}]{comment:scale}
\bibinfo{note}{A finite-size scaling of the data for $p = 0$ using the known
  exponents $\nu = 3/4$ ($\alpha = 1/2$) shows that corrections to scaling are
  small.}

\bibitem[{\citenamefont{Geyer}(1991)}]{geyer:91}
\bibinfo{author}{\bibfnamefont{C.}~\bibnamefont{Geyer}}, in
  \emph{\bibinfo{booktitle}{23rd Symposium on the Interface}}, edited by
  \bibinfo{editor}{\bibfnamefont{E.~M.} \bibnamefont{Keramidas}}
  (\bibinfo{publisher}{Interface Foundation}, \bibinfo{address}{Fairfax
  Station}, \bibinfo{year}{1991}), p. \bibinfo{pages}{156}.

\bibitem[{\citenamefont{Hukushima and Nemoto}(1996)}]{hukushima:96}
\bibinfo{author}{\bibfnamefont{K.}~\bibnamefont{Hukushima}} \bibnamefont{and}
  \bibinfo{author}{\bibfnamefont{K.}~\bibnamefont{Nemoto}},
  \emph{\bibinfo{title}{Exchange {M}onte {C}arlo method and application to spin
  glass simulations}}, \bibinfo{journal}{J. Phys. Soc. Jpn.}
  \textbf{\bibinfo{volume}{65}}, \bibinfo{pages}{1604} (\bibinfo{year}{1996}).

\bibitem[{com({\natexlab{c}})}]{comment:ising}
\bibinfo{note}{For the two-dimensional Ising ferromagnet on the square lattice
  $\nu = 1$ and $\alpha = 0$. Note that for the three-body Ising model on the
  triangular lattice $\nu = \alpha = 2/3$ \cite{baxter:82}.}

\bibitem[{\citenamefont{Baxter}(1982)}]{baxter:82}
\bibinfo{author}{\bibfnamefont{R.}~\bibnamefont{Baxter}},
  \emph{\bibinfo{title}{{Exactly Solved Models in Statistical Mechanics}}}
  (\bibinfo{publisher}{Academic Press}, \bibinfo{address}{London},
  \bibinfo{year}{1982}).

\end{thebibliography}

\end{document}